\documentclass[aps,prb,twocolumn,superscriptaddress,amsmath,amssymb]{revtex4-2}

\usepackage{graphicx}% Include figure files
\usepackage{dcolumn}% Align table columns on decimal point
\usepackage{bm}% bold math
\usepackage[hidelinks]{hyperref}
\usepackage{subcaption}
\usepackage[abs]{overpic}  % to place text over images
\usepackage[justification=raggedright,singlelinecheck=false]{caption} % caption aligned left 
\usepackage{booktabs} % for table lines
\usepackage{upgreek} % \upmu possible

\begin{document}

\title{Hybrid Quantum Systems: Coupling Single-Molecule Magnet Qudits with\\Industrial Silicon Spin Qubits}

\author{Daniel Schroller}
\email{daniel.schroller@kit.edu}
\affiliation{Physikalisches Institut, Karlsruhe Institute of Technology (KIT), Karlsruhe, Germany}

\author{Daniel Sitter}
\affiliation{Physikalisches Institut, Karlsruhe Institute of Technology (KIT), Karlsruhe, Germany}

\author{Thomas Koch}
\affiliation{Physikalisches Institut, Karlsruhe Institute of Technology (KIT), Karlsruhe, Germany}

\author{Viktor Adam}
\affiliation{Institute for Quantum Materials and Technologies, Karlsruhe Institute of Technology (KIT), Karlsruhe, Germany}

\author{Noah Glaeser}
\affiliation{Physikalisches Institut, Karlsruhe Institute of Technology (KIT), Karlsruhe, Germany}

\author{Clement~Godfrin}
\affiliation{Interuniversity Microelectronics Centre (imec), Leuven, Belgium}

\author{Stefan Kubicek}
\affiliation{Interuniversity Microelectronics Centre (imec), Leuven, Belgium}

\author{Julien Jussot}
\affiliation{Interuniversity Microelectronics Centre (imec), Leuven, Belgium}

\author{Roger Loo}
\affiliation{Interuniversity Microelectronics Centre (imec), Leuven, Belgium}
\affiliation{Department of Solid-State Sciences, Ghent University, Ghent, Belgium}

\author{Yosuke Shimura}
\affiliation{Interuniversity Microelectronics Centre (imec), Leuven, Belgium}

\author{Danny~Wan}
\affiliation{Interuniversity Microelectronics Centre (imec), Leuven, Belgium}

\author{Yaorong Chen}
\affiliation{Institute of Nanotechnology, Karlsruhe Institute of Technology (KIT), Karlsruhe, Germany}

\author{Mario Ruben}
\affiliation{Institute for Quantum Materials and Technologies, Karlsruhe Institute of Technology (KIT), Karlsruhe, Germany}
\affiliation{Institute of Nanotechnology, Karlsruhe Institute of Technology (KIT), Karlsruhe, Germany}

\author{Kristiaan De Greve}
\affiliation{Interuniversity Microelectronics Centre (imec), Leuven, Belgium}
\affiliation{Department of Electrical Engineering, ESAT-MNS and Proximus Chair in Quantum Science and Technology, KU Leuven, Leuven, Belgium}

\author{Wolfgang Wernsdorfer}
\email{wolfgang.wernsdorfer@kit.edu}
\affiliation{Physikalisches Institut, Karlsruhe Institute of Technology (KIT), Karlsruhe, Germany}
\affiliation{Institute for Quantum Materials and Technologies, Karlsruhe Institute of Technology (KIT), Karlsruhe, Germany}

\date{\today}

\begin{abstract}
Molecular spin qudits offer an attractive platform for quantum memory, combining long coherence times with rich multi-level spin structures. Terbium bis(phthalocyaninato) (TbPc$_2$) exemplifies such systems, with demonstrated quantum control and chemical reproducibility. In hybrid quantum architectures, TbPc$_2$ can act as the primary memory element, with semiconductor qubits providing scalable readout and coupling. Here we present a step toward such a hybrid system: using an industrially manufactured silicon metal-oxide-semiconductor (SiMOS) spin qubit to detect electronic spin transitions of an ensemble of TbPc$_2$ molecules. The readout is based on a compact and robust protocol that applies a microwave pulse while all gate voltages defining the qubit are held at a fixed operating point.
This protocol, which combines simultaneous Rapid adiabatic Passage and Spin-Selective tunneling (RPSS), enables high-contrast resonance detection and avoids repeated \mbox{\(\pi\)-pulse} recalibration common in decoupling schemes. By demonstrating ensemble detection, we establish a foundation for integrating molecular quantum memories with industrial qubit platforms and mark an important step toward single-molecule hybrid quantum technologies.%
\end{abstract}

\maketitle

\section*{Introduction}
Quantum information processing stands to benefit from architectures that combine long-lived quantum states with the integrability of solid-state platforms. Among the many systems under development, single-molecule magnets (SMMs) have long been regarded as promising candidates for quantum memory and logic, owing to their compactness, the ability to produce chemically identical systems with exact reproducibility~\cite{Atzori2019, Wernsdorfer2019, MorenoPineda2024}, chemical tunability~\cite{Atzori2019, Wernsdorfer2019}, and inherently multilevel spin structure~\cite{MorenoPineda2018, MorenoPineda2024}. These molecular spin qudits can occupy more than two quantum states simultaneously, enabling more efficient information encoding and manipulation~\cite{Jankovic2024}. A prototypical example is terbium bis(phthalocyaninato) (TbPc$_2$)~\cite{Ishikawa2003}, which exists in both neutral and anionic forms. Most demonstrations of single-molecule quantum control have used the neutral radical. Both charge states host coupled electronic and nuclear spin degrees of freedom, giving rise to a rich spectrum of magnetically addressable levels. 
Benefiting from the shielding of the 4f electrons, TbPc$_2$ shows remarkably long-lived nuclear spin dynamics: the spin-lattice relaxation time is \mbox{$T_1 = 10-30\,\mathrm{s}$}~\cite{Thiele2013,Thiele2014}, and the inhomogeneous dephasing time is $T_2^* \approx 200\,\upmu\mathrm{s}$~\cite{Godfrin2017}. These coherence properties, together with atomic-level chemical reproducibility, make TbPc$_2$ an attractive candidate for molecular-based quantum information processing. Coherent quantum control has been achieved at the single-molecule level, including the realization of multi-level algorithms such as Grover’s algorithm~\cite{Godfrin2017} and the realization of an iSWAP gate~\cite{Godfrin2018}.

% Legacy readout techniques
Despite these advantages, the very shielding of the electronic states that grants exceptional coherence also renders direct readout and circuit-level integration difficult. Previous approaches have achieved spin-state readout by incorporating TbPc$_2$ molecules into nanoelectronic transport structures, such as carbon nanotubes~\cite{Urdampilleta2011, Ganzhorn2013} and gold break junctions~\cite{Vincent2012, Thiele2014, Godfrin2017, Lumetti2016}, where the magnetization dynamics modulate conductance. While elegant and informative, such readout methods face significant limitations in scalability and compatibility with established semiconductor-based quantum device technologies.

To address these limitations, we demonstrate a hybrid quantum architecture that couples a TbPc$_2$ molecular qudit to an industrial silicon metal-oxide-semiconductor (SiMOS) spin qubit. In this scheme, TbPc$_2$ provides the primary quantum element for information storage and manipulation, whereas the SiMOS qubit furnishes a robust, cryo-compatible, CMOS-integrated readout interface and a prospective mediator for qudit-qudit interactions. SiMOS quantum dots combine wafer-scale manufacturability~\cite{Jock2018,Steinacker2024} with mature spin-to-charge conversion readout schemes~\cite{Elzerman2004,Hu2021}. On-chip microwave lines are routinely integrated in SiMOS~\cite{Veldhorst2015} and can be repurposed to drive molecular spin transitions, further underscoring SiMOS as a practical interface to molecular quantum systems.

Crucially, the two platforms are complementary. TbPc$_2$ offers long-lived spin dynamics, with nuclear-spin dephasing times of $T_2^*\!\approx\!200\,\upmu \mathrm{s}$ reported at the single-molecule level~\cite{Godfrin2017}, and in our ensemble measurements (see Results) we observe an electronic-spin relaxation time of $T_1 \approx 107 \,\mathrm{min}$. By contrast, isotopically enriched $^{28}$Si SiMOS quantum-dot spins typically display much shorter dephasing times, with $T_2^*$ in the few $\upmu \mathrm{s}$ under comparable conditions~\cite{Zhao2019, Veldhorst2015, Vahapoglu2022}, and electronic-spin $T_1$ values up to the $10\,\mathrm{s}$ range~\cite{Stano2022}. This contrast motivates using the molecular spin as a long-lived quantum memory, while the SiMOS qubit supplies fast, integrated readout and a pathway to mediate couplings between molecular qudits.

% Experimental demonstarion of electron spin coupling, as a way to single TbPc2 molecule sensing
Here we present a hybrid quantum system in which a SiMOS spin qubit is used to read out the electronic spin state of an ensemble of [TbPc$_2]^{-}$ molecules deposited atop the qubit device. In this charge state, the radical of the ligand is absent, suppressing radical-mediated coupling while preserving the molecular spin-qudit structure. The experiment employs a compact sensing protocol combining spin-selective tunneling and rapid adiabatic passage (RAP) in a single microwave burst and avoids repeated \(\pi\)-pulse recalibration common in decoupling schemes.
Throughout each acquisition, all qubit gate voltages are held fixed at a single, stable bias point; only the microwave drive is varied. Qubit resonance shifts encode the spin state of the [TbPc$_2$]$^{-}$ ensemble. We observe magnetic hysteresis, angular-dependent anisotropy, and thermally activated relaxation, which are canonical signatures of Tb-based SMM magnetization dynamics.

The present implementation is limited by the vertical spacing between the spin qubit and the molecular layer, a constraint that can be alleviated with chip layouts engineered specifically for hybrid operation. By employing molecular ensembles to boost the dipolar signal at the qubit, we realize a working proof of concept. This hybrid architecture couples a chemically engineered molecular spin qudit to an industrial silicon spin qubit, marking a key step toward scalable, high-coherence quantum memory embedded in semiconductor platforms.

~\\
\section*{Methods}
\paragraph*{\textbf{Hybrid Quantum Device Architecture.}}
\label{subsec:Device}
\begin{figure}[t]
    \centering
    \includegraphics[width=0.9\columnwidth]{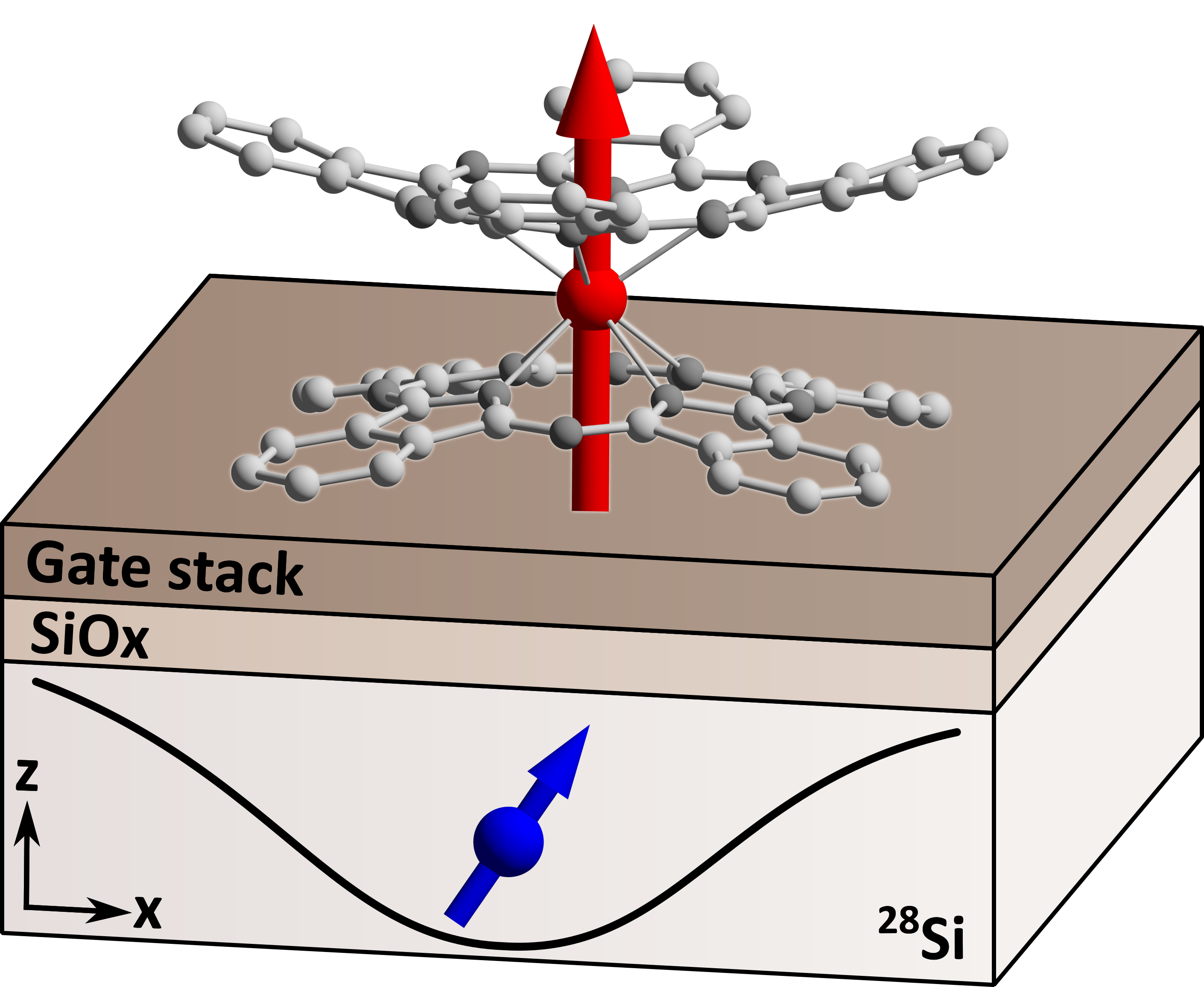}  
    \caption{\textbf{Hybrid system combining a single-molecule magnet and a silicon spin qubit.}
    A single [TbPc$_2$]$^{-}$ molecule is placed above a silicon quantum-dot device and is magnetically coupled to the qubit via its dipolar field. The Tb$^{3+}$ ion is depicted in red with its magnetic moment aligned with the molecular easy axis and indicated by a red arrow. The quantum-dot electron (blue) resides in isotopically enriched $^{28}$Si within a lateral confinement potential (black) defined by the metallic gate stack (brown), which is separated from the silicon by an SiO$_x$ layer. The gate stack and oxide together set the vertical separation between the molecular spin and the semiconductor qubit. Schematic not to scale; the quantum dot is an order of magnitude larger than the SMM}.
    \label{fig:Schematic_coupling}
\end{figure}

The hybrid device developed in this work integrates a molecular spin qudit with a semiconductor spin qubit, combining chemically engineered quantum memory with solid-state readout. The central quantum element is the single-molecule magnet [TbPc$_2]^{-}$, which serves as the quantum memory. In Tb$^{3+}$, strong spin-orbit coupling and the axial ligand field stabilize a ground state doublet with predominant Ising character ($J_z=\pm6$), isolated from the first excited doublet ($J_z=\pm5$) by an anisotropy barrier of approximately 600\,K~\cite{Ishikawa2003, MorenoPineda2024}, effectively freezing the spin orientation at cryogenic temperatures. Hyperfine coupling to the $I = 3/2$ nuclear spin splits the ground state manifold into four unequally spaced levels, forming a robust four-level qudit structure.

These levels can be coherently controlled and exhibit long intrinsic coherence, with measured Ramsey coherence times of $T_2^* \approx 200~\upmu$s~\cite{Godfrin2017}. Nuclear spin-state readout is performed indirectly by determining the magnetic field at which the electronic spin flips. Owing to the hyperfine interaction, this switching field depends on the nuclear spin projection, enabling spin-state discrimination through magnetization dynamics~\cite{Thiele2013}. A detailed depiction of the [TbPc$_2]^{-}$ energy-level structure is provided in Refs.~\cite{Vincent2012, MorenoPineda2024}.

To detect these spin transitions electrically, we interface the molecule with a silicon metal-oxide-semiconductor (SiMOS) spin qubit (Fig.~\ref{fig:Schematic_coupling}), whose Larmor frequency is highly sensitive to local magnetic fields. This pairing combines the coherence and multilevel encoding of the molecular qudit with the high-fidelity, scalable readout capabilities of a semiconductor spin qubit.

Among state-of-the-art platforms, SiMOS and SiGe architectures dominate silicon qubit development. For surface-coupled sensing applications, their vertical geometries are critically distinct. Si/SiGe devices often incorporate thick cap and buffer layers exceeding 40\,nm~\cite{Pena2024, Koch2025, vanRiggelen-Doelman2024}, limiting proximity to external spins. To minimize this separation, and thereby enhance sensor-sample coupling and improve signal strength, we employ a SiMOS platform. 

Additional advantages of the SiMOS platform include the absence of micromagnets that might distort local fields and the integration of a microwave transmission line. Originally designed for coherent spin control~\cite{Koppens2006}, this line could enable direct driving of the molecular spin transitions. Fabrication details of the industrial SiMOS chip are provided in Appendix~\ref{sec:fabrication_details}.

\begin{figure}[tb]
    \centering
    \includegraphics[width=\columnwidth]{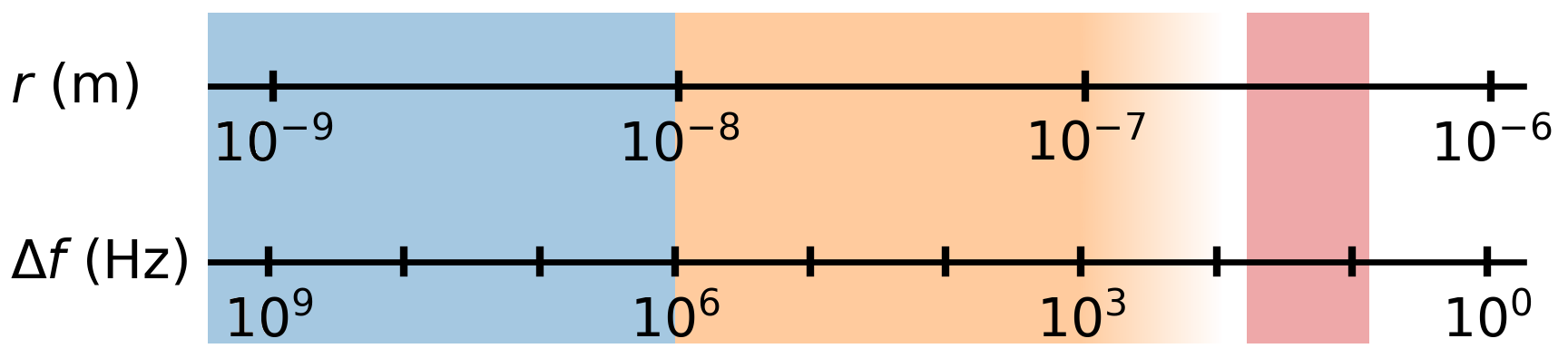}
    \caption{\textbf{Distance dependence of dipolar coupling in the hybrid device.}
    The two axes show how the vertical separation \(r\) between a single [TbPc$_2]^{-}$ molecule and a silicon spin qubit maps onto the corresponding qubit frequency shift \(\Delta f\). In the current industrial device, the gate stack and passivation layer set a separation of \(r \approx 250\,\mathrm{nm}\) (red region), leading to a corresponding frequency shift \(\Delta f \approx  60\,\mathrm{Hz}\). This is below the typical detection limit of dynamical decoupling techniques (orange region, assuming $T_{2} \sim 100\,\upmu\mathrm{s}$). To amplify the signal, a dilute ensemble of molecules is used, increasing the magnetic field by several orders of magnitude and shifting the qubit response into the sensitive regime of the simultaneous rapid adiabatic passage and spin-selective tunneling (RPSS) protocol (blue region).
    }
    \label{fig:axis}
\end{figure}

The magnetic coupling between the [TbPc$_2$]$^{-}$ molecule and the silicon spin qubit is set primarily by their vertical separation. In the present device, which was not specifically engineered for hybrid operation, the gate stack and passivation define a spacing of \(r \approx 250\,\mathrm{nm}\). At this distance, the on-axis dipolar field of a Tb$^{3+}$ moment \mbox{\(m=g_J J \mu_{\mathrm B}\approx 9\mu_{\mathrm B}\)}~\cite{Kittel2005} yields \mbox{\(B = \mu_0 m/(2\pi r^3) \approx 1\,\mathrm{nT}\)}~\cite{Jackson1999}. This corresponds to an electron-spin Larmor shift \mbox{\(\Delta f = 2 \gamma_e B \approx 60\,\mathrm{Hz}\)} \mbox{(\(\gamma_e \approx 28\,\mathrm{GHz/T}\)}~\cite{Hwang2017}). With a Hahn-echo coherence time \(T_2 \sim 100\,\upmu\mathrm{s}\), the Fourier-limited (Lorentzian) linewidth is $1/(\pi \, T_2) \approx 3\,\mathrm{kHz}$~\cite{abragam1961}; a single-molecule shift in $\Delta f$ is therefore well below our detectable threshold (Fig.~\ref{fig:axis}).

To obtain a measurable response, we use a highly diluted ensemble of [TbPc$_2$]$^{-}$ molecules (Appendix~\ref{subsec:TbPc2}), in which the molecular spins are sufficiently separated to suppress intermolecular coupling and preserve single-molecule behavior. The collective dipolar field at the qubit then enhances the net frequency shift to the tens-of-megahertz range. Detection is performed with the RPSS protocol described below. This approach constitutes a first experimental step toward hybrid integration of magnetic molecules within an industrial SiMOS platform.

\paragraph*{\textbf{Spin resonance detection by simultaneous rapid adiabatic passage and spin-selective tunneling.}}
\label{subsec:resonance_detection}
\begin{figure}[t] 
  \centering
  \includegraphics[width=\columnwidth]{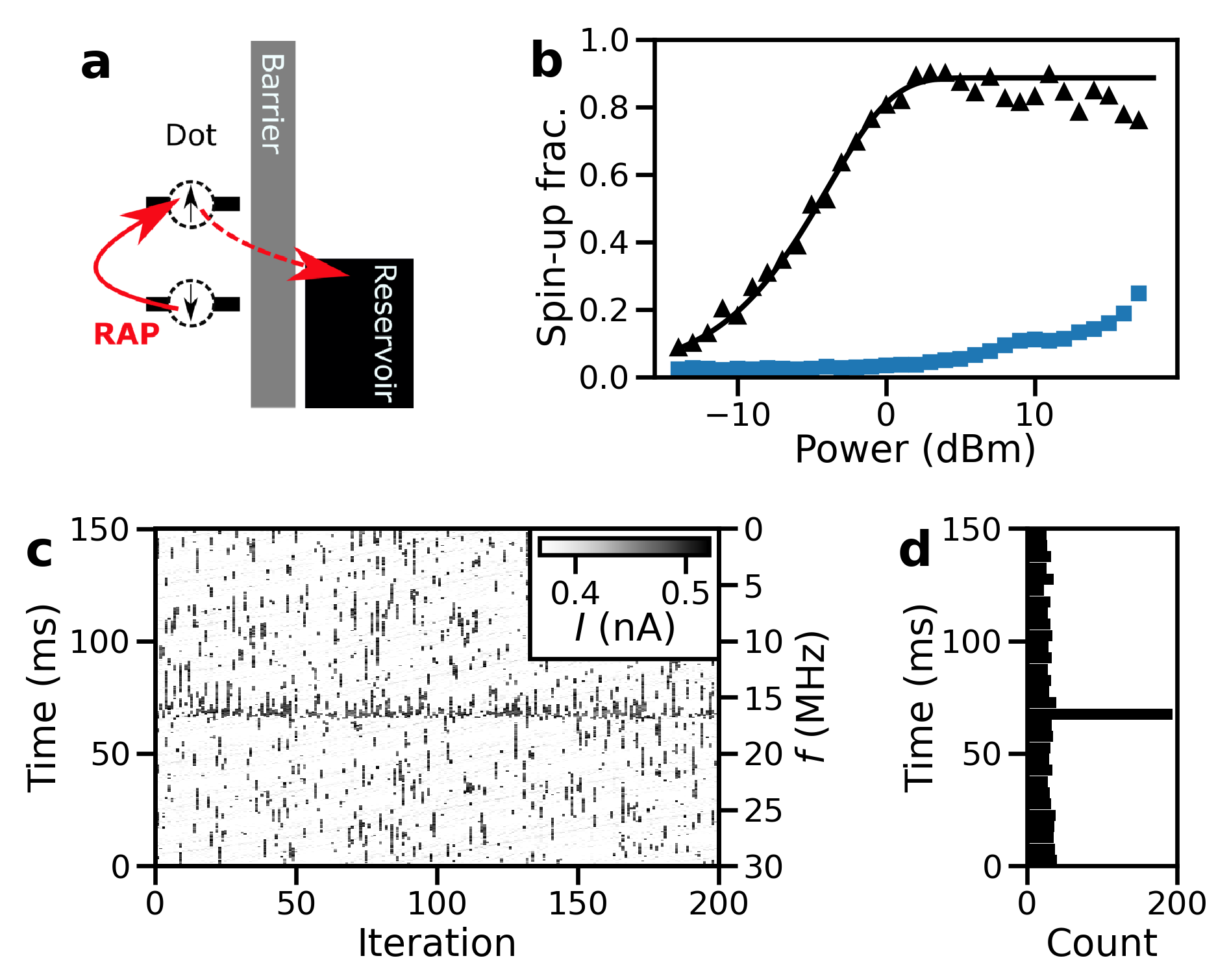}
    \caption{\textbf{Spin resonance detection via RPSS protocol.} 
    \textbf{a} A frequency-chirped microwave pulse drives the spin-up transition, triggering spin-dependent tunneling into the reservoir for charge detection. 
    \textbf{b}~Resonance visibility (black) and background (blue) as a function of applied microwave power at a chirp speed of $0.2\,\mathrm{GHz/s}$. The solid line is a fit to the Landau-Zener model (Appendix~\ref{sec:appendix_RPSS}) for power levels below 5\,dBm, yielding a maximum visibility of $89 \pm 1\%$. At higher powers, thermally activated background events reduce the observed visibility.
    \textbf{c}~Time-resolved data from 200 single-shot traces during a 150\,ms chirp over 30\,MHz at 8\,dBm drive power. Tunneling events appear as discrete “blips” in the sensor current. The frequency axis is offset by  18.57\,GHz.
    \textbf{d}~Histogram of blip onset times extracted from \textbf{c}, showing a pronounced resonance peak corresponding to the qubit’s Larmor frequency.}    
     \label{fig:Methods}
\end{figure}

To detect changes in the local magnetic field caused by the [TbPc$_2]^{-}$ ensemble, we monitor the spin resonance frequency of a SiMOS spin qubit $f_0 = g\mu_\mathrm{B}B/h$. In our device, the molecular field shifts $f_0$ by up to 34\,MHz, requiring a robust and broadband detection protocol.

We employ a rapid adiabatic passage (RAP) technique, in which a frequency-chirped microwave pulse drives the qubit through resonance, resulting in spin inversion. Adiabatic transfer occurs when the chirp rate is sufficiently slow relative to the square of the Rabi frequency (the Landau-Zener adiabaticity condition), so that the inversion probability approaches unity~\cite{abragam1961,Harris1972,Li2023}. In contrast to fixed-frequency $\pi$-pulse protocols, which require finely stepped scans and repeated recalibration, RAP achieves robust, high-contrast inversion without precise tuning of the pulse length or center frequency, provided the chirp spans the resonance and the adiabatic criterion is met.

To convert spin transitions into electrical signals, we operate the qubit in the Elzerman spin-selective readout regime~\cite{Elzerman2004}: a spin-down electron remains confined in the dot, while an excited spin-up electron tunnels into the reservoir (Fig.~\ref{fig:Methods}\textbf{a}). These tunneling events are detected in real time as current steps, or “blips,” using a proximal SET.

Combining RAP with this spin-dependent tunneling yields the simultaneous rapid adiabatic passage and spin-selective tunneling (RPSS) protocol. The entire measurement consists of a single chirped microwave pulse while maintaining all qubit gate voltages at a single, stable bias point. Each tunneling event is timestamped, allowing frequency shifts to be directly mapped to changes in blip onset time (Fig.~\ref{fig:Methods}\textbf{c,d}).

The resonance visibility depends on the interplay of chirp rate, microwave power, and tunnel coupling. In calibration experiments, we used 150\,ms chirps over 30\,MHz at 8\,dBm power (setup details in Appendix~\ref{subsec:setup}) with a tunnel rate of ~300\,Hz. Under these conditions the visibility peaked at 89\% for moderate drive powers (Fig.~\ref{fig:Methods}\textbf{b}). At higher powers, however, microwave heating increases thermally activated off-resonant tunneling that empties the dot before the chirp reaches resonance. Given the relatively weak tunnel coupling, the dot then remains empty when the sweep crosses \(f_0\), so no RAP-induced tunneling can occur. These premature escape events induce a sensor dead time and reduce the measured visibility.

For quantitative detection, we compute a histogram of tunneling-event start times aggregated over multiple RPSS measurements. A 1\,MHz bin width ensures the resonance is well-captured, and visibility is defined as the difference between spin-up probabilities on- and off-resonance (Fig.~\ref{fig:Methods}\textbf{d}). Further details, including Rabi frequency extraction are provided in Appendix~\ref{sec:appendix_RPSS}.

For the magnetometry measurements presented in the Results section, we expanded the chirp to \(100\,\mathrm{MHz}\) to capture the full hysteresis loop. With the pulse length fixed at \(150\,\mathrm{ms}\), the resulting sweep rate exceeds the adiabatic threshold and reduces visibility to below 50\%. Nevertheless, the contrast remained sufficient to localize the qubit resonance.

A key advantage of RPSS over conventional RAP combined with sequential readout is the reduction in total measurement time. In our implementation, each single-shot RAP trace lasts 5\,ms (limited by DC blip detection, which requires reservoir coupling rates of \(\sim\!1\,\mathrm{kHz}\)). Using sequential narrow-band RAP and separate readout over 100 frequency windows would increase total duration more than threefold. 

Each full resonance measurement (50 averages) takes approximately 15\,s. With a frequency resolution of 1\,MHz and chirp span of 100\,MHz, this yields a field sensitivity of 138\,$\upmu$T$/\sqrt{\mathrm{Hz}}$, sufficient to track the dynamic magnetic response of the [TbPc$_2]^{-}$ ensemble.

\section*{Results}
\label{sec:Results}
\begin{figure*}[t]
  \centering
  \includegraphics[width=\textwidth]{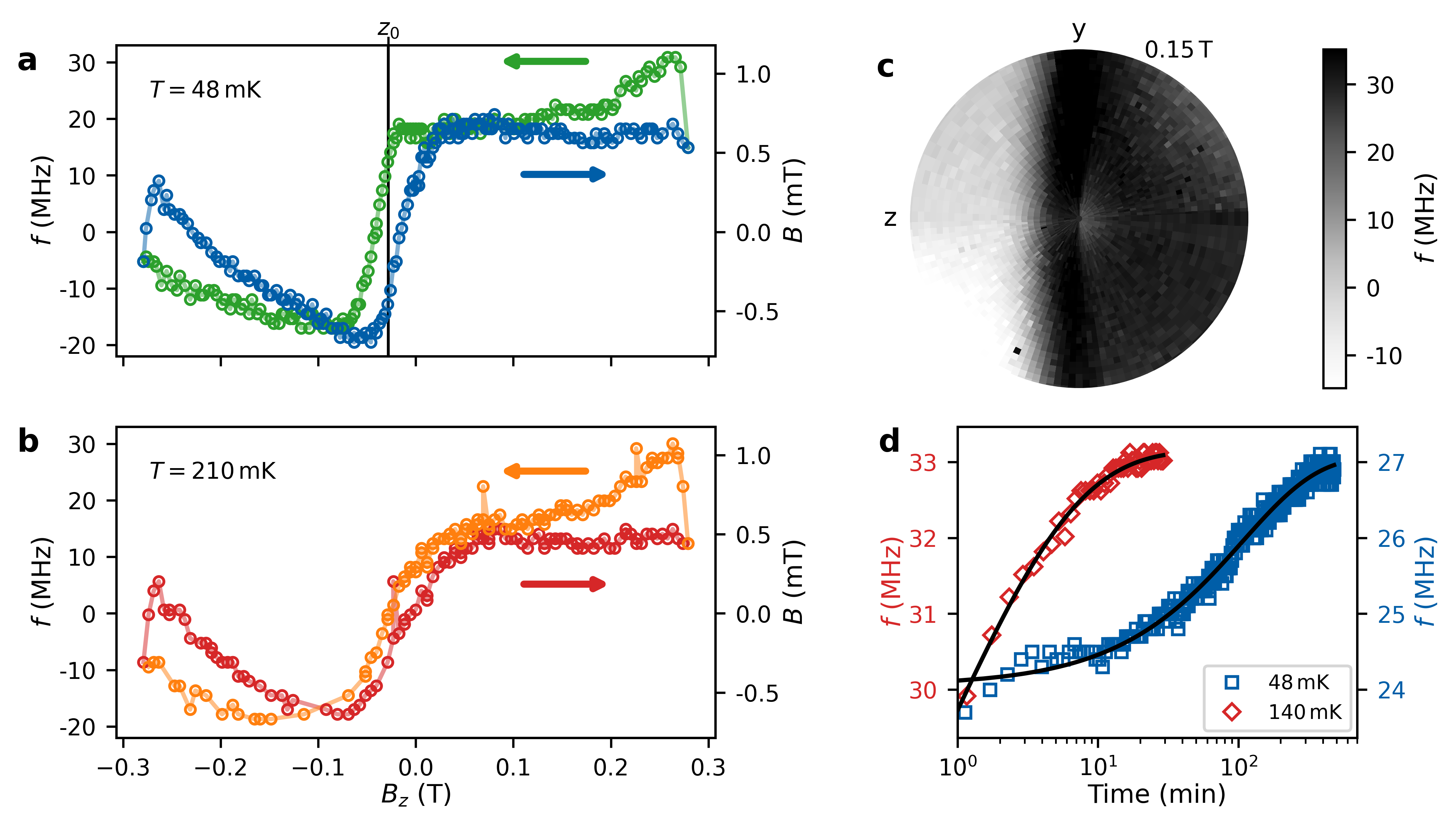}
  \caption{
    \textbf{Silicon spin qubit-based magnetometry of a [TbPc$_2]^{-}$ ensemble.}  
    Qubit resonance frequency (offset by 18.435\,GHz) as a function of the magnetic field component along the [TbPc$_2]^{-}$ easy axis ($z$-axis), measured during angular sweeps of the external field in the $xz$-plane at 48\,mK (\textbf{a}) and 210\,mK (\textbf{b}). Each data point represents 50 single-shot measurements and a 1\,MHz bin width. A $g$-factor of 2~\cite{Fogarty2018} is used to convert the qubit frequency to the corresponding magnetic field, up to a constant offset. Forward and backward sweeps reveal pronounced magnetic hysteresis associated with a full spin-state reversal of the [TbPc$_2]^{-}$ ensemble ($J = \pm6$). Artifacts from the 3D vector magnet appear at the sweep reversal points ($\pm$0.280\,T).
    \textbf{c} Polar plot of the resonance frequency at 48\,mK during angular sweeps in the $yz$-plane (radial axis spans $0.00\, \mathrm{T}$ to $0.15\,\mathrm{T}$; sweep direction from center outward). A sharp frequency transition along the $z$-axis confirms alignment with the [TbPc$_2]^{-}$ easy axis.
    \textbf{d} Time-resolved relaxation of the [TbPc$_2]^{-}$ ensemble after saturating at $-0.280$\,T. The field is rapidly swept to the setpoint labeled $z_0$ in \textbf{a}, after which the qubit’s resonance frequency is monitored over time. A small drift of the applied vector field during the sequence shifted the effective $z_0$,  reducing  the frequency visibility and slightly changing its absolute value. Each data point is the average of 100 acquisitions (30\,s per point) taken with a chirp rate of 0.267\,GHz/s and a frequency binning of 0.1\,MHz. The data are fitted with stretched exponentials (see main text), yielding relaxation times $\tau=(107\pm2)\,\mathrm{min}$ and $\tau=(0.8\pm0.3)\,\mathrm{min}$ at 48\,mK and 140\,mK, respectively.
}
  \label{fig:Results}
\end{figure*}

We employ a SiMOS spin qubit as a cryogenic quantum sensor to investigate the magnetization dynamics of a [TbPc$_2]^{-}$ ensemble (for details refer to Appendix~\ref{subsec:TbPc2}). The qubit, characterized with a relaxation time of \mbox{\(T_1 = 110 \pm 7\,\mathrm{ms}\)} and a Ramsey coherence time of \mbox{\(T_2^* = 3.1 \pm 0.1\,\upmu\mathrm{s}\)} (see Appendix~\ref{sec:T1_T2}), enables sensitive detection of local magnetic fields via the RPSS protocol introduced earlier. 

As a benchmark for the sensing platform, we probe key magnetic signatures of the SMM ensemble, including magnetic hysteresis~\cite{Vincent2012}, angular-dependent anisotropy~\cite{Perfetti2016}, and thermally activated spin relaxation~\cite{Ishikawa2003}. A static magnetic field of \(660\,\mathrm{mT}\) is applied to establish robust Zeeman splitting for reliable qubit control and readout.Throughout the experiments, the field magnitude is held fixed, and only its orientation is varied, which we set using a three-dimensional vector magnet (Appendix~\ref{subsec:magnet_calib}).

\paragraph*{\textbf{Magnetic hysteresis.}}
To characterize the molecular magnet, we probe how the response along its easy axis varies with field angle. Shown in Fig.~\ref{fig:Results}\textbf{a} is the qubit resonance frequency as a function of the $z$-component of the applied field which is aligned with the [TbPc$_2$]$^{-}$ easy axis. The data were obtained during angular sweeps in the $xz$-plane at 48\,mK, for both trace and retrace.
A pronounced hysteresis loop is observed, with the loop center offset by $-28.3$\,mT due to residual $z$-contributions from the $x$-coil of the vector magnet. A remanent magnetization of 0.4\,mT and a coercive field of 12.8\,mT are extracted from the data.

At an elevated temperature of 210\,mK (Fig.~\ref{fig:Results}\textbf{b}), thermally enhanced electronic spin reversal within the Ising ground doublet leads to a broadened hysteresis loop, as indicated by reduced remanence and smoother transitions in the resonance shift.

A slight non-orthogonality of the vector-magnet coil axes, compounded by suboptimal correction angles, results in a loop-center offset of $-28.3$\,mT and a weak slope for negative $B_z$. In both data sets, reproducible features appear at the sweep extrema ($\pm 280\,\mathrm{mT}$) upon reversal of the sweep direction: the field increases by $\sim 0.5\,\mathrm{mT}$ after each reversal, producing a corresponding increase in the resonance frequency. We attribute these features to the rearrangement of magnetic-flux vortices in the NbTi type-II superconducting vector magnet~\cite{Nabialek2010}. As the magnet is thermally anchored to the 4\,K stage, these artifacts are temperature independent and not sample related.

Control measurements without the [TbPc$_2]^{-}$ crystal (Appendix~\ref{sec:hysteresis_no_SMM}) confirm that the end-of-sweep artifacts persist while the central hysteresis loop disappears, supporting their origin in the superconducting magnet rather than in the molecular ensemble.

\paragraph*{\textbf{Magnetic anisotropy.}}
To further verify the alignment of the [TbPc$_2]^{-}$ crystal’s easy axis, we perform an angular-resolved magnetometry scan with the magnetic field projected in the $yz$-plane. The resulting data are shown in Fig.~\ref{fig:Results}\textbf{c}.

A sharp resonance frequency shift is observed along the $z$-axis, indicating that the easy axis of the [TbPc$_2]^{-}$ ensemble is aligned with the $z$-direction of the 3D vector magnet. This angular dependence confirms both the pronounced magnetic anisotropy of the molecule and its dipolar contribution to the local magnetic environment sensed by the qubit.

\paragraph*{\textbf{Thermal relaxation.}}
A characteristic feature of single-molecule magnets is the strong temperature dependence of their magnetization relaxation~\cite{Ishikawa2003}, reflecting thermally activated reversal over the anisotropy barrier. To reveal this behavior, we track the time evolution of the [TbPc$_2$]$^{-}$ ensemble’s electronic magnetization via the qubit resonance.
The sample is first saturated at $-0.280\,\mathrm{T}$ along $z$, then returned to zero field in $z$ (see $z_0$ in Fig.~\ref{fig:Results}\textbf{a}). The qubit resonance is monitored over time at $48\,\mathrm{mK}$ and $140\,\mathrm{mK}$ (Fig.~\ref{fig:Results}\textbf{d}). At both temperatures, a resonance shift of approximately 3\,MHz is observed, indicative of a time-dependent magnetization change in the [TbPc$_2]^{-}$ ensemble. The relaxation proceeds significantly faster at elevated temperature, consistent with thermally activated dynamics.

The data are fitted to a stretched exponential function \mbox{$f(t) = A \cdot \exp(-\left(t/\tau\right)^\beta) + C,$}
where $A$ is the amplitude, $\tau$ the characteristic relaxation time, $\beta$ the stretch exponent, and $C$ an offset. This functional form captures the distributed relaxation behavior characteristic of SMM ensembles~\cite{Edholm2000}, where variations in local environments result in a broad spectrum of relaxation rates.

At 48\,mK, the extracted relaxation time is \mbox{$\tau = (107 \pm 2)$\,min} with a stretching exponent \mbox{$\beta = 0.81\pm 0.02$}, indicating moderately distributed relaxation dynamics. At 140\,mK, the relaxation accelerates markedly, with \mbox{$\tau = (0.8\pm0.3)$\,min} and a reduced exponent \mbox{$\beta = 0.46\pm0.06$}. This decrease in $\beta$ reflects a broader spread of relaxation times at elevated temperatures, consistent with thermally activated access to a wider range of microscopic relaxation pathways~\cite{Edholm2000, Woodruff2013}. These measurements demonstrate that the SiMOS qubit can track the magnetization dynamics of a [TbPc$_2]^{-}$ ensemble over timescales spanning minutes to hours, establishing a robust foundation for more advanced hybrid quantum sensing schemes.

\section*{Discussion and Outlook}
In this work, we have demonstrated the feasibility of a hybrid quantum sensing architecture that combines a silicon metal-oxide-semiconductor (SiMOS) spin qubit with a molecular spin ensemble, using terbium bis(phthalocyaninato) ([TbPc$_2]^{-}$) as the quantum memory element. The SiMOS qubit acts as a sensitive and CMOS-compatible interface for readout, enabling the detection of subtle magnetic resonance shifts induced by the ensemble’s magnetization dynamics. Our compact simultaneous rapid adiabatic passage and spin-selective tunneling (RPSS) protocol provides robust resonance detection without the need for dynamic recalibration or feedback, operating efficiently at a constant working point.

The present device successfully resolves characteristic signatures of the electronic spin states of the molecular ensemble. These include pronounced hysteresis loops, clear angular anisotropy aligned with the molecular easy axis, and thermally activated relaxation dynamics with long characteristic times exceeding 100 minutes at cryogenic temperatures. 

A primary limitation of the current device arises from the substantial vertical separation ($250\,\mathrm{nm}$) between the qubit and the molecular qudit, significantly reducing magnetic coupling and limiting sensitivity. Future device iterations, involving optimized gate stacks and fabrication processes, aim to reduce this separation to below 10\,nm, potentially enabling detection of single [TbPc$_2]^{-}$ molecules. Additionally, precise placement of individual single-molecule magnets remains a challenging but crucial task.

To significantly enhance detection sensitivity and spectral precision, we will employ dynamical decoupling sequences. By refocusing low-frequency noise, these protocols suppress decoherence and sharpen the frequency response of the measurement, yielding narrow, high-contrast interference fringes. Simulations (Appendix Fig.~\ref{fig:sim_pulses}) indicate substantial gains in frequency resolution and magnetic sensitivity. Such improvements are pivotal for resolving the subtle field shifts produced by individual molecular spin transitions, enabling a transition from ensemble averages to single-molecule readout.

Ultimately, our work highlights a promising route toward hybrid quantum platforms that exploit the chemically identical reproducibility of SMMs together with the scalability, compact footprint, and CMOS compatibility inherent in SiMOS-based qubits. Leveraging SiMOS architectures that support dense qubit arrays and laterally mobile sensor qubits~\cite{Tosato2025,DeSmet2025} significantly expands possibilities for coupling and sensing configurations. Continued device optimization, combined with sophisticated quantum pulse control, positions SiMOS-based hybrid quantum systems as versatile tools for quantum computing, quantum sensing, and fundamental studies of spin phenomena in molecular and solid-state materials.

\begin{acknowledgments}
We thank C. Sürgers for valuable advice and support with the measurement electronics, T. Cubaynes for designing the sample printed circuit board, and E.~Moreno-Pineda, S.~Paul, and A.~Sunil for $\upmu$SQUID precharacterization of the [TbPc$_2]^{-}$ crystals. We acknowledge financial support from the German Research Foundation (DFG) through the Gottfried Wilhelm Leibniz Award (ZVN-2020 WE 4458-5). The work at imec was supported by the imec Industrial Affiliation Programme on Quantum Computing and the European Union’s Horizon 2020 Research and Innovation Programme under grant agreement No. 101174557 (QLSI2).
\end{acknowledgments}

\section*{Author contributions}
D.Sc. wrote the manuscript with input from all co-authors. D.Sc. and D.Si. performed the experiments and analyzed the data. N.G. developed parts of the data-analysis code. S.K. contributed as integration engineer, J.J. was responsible for electron beam lithography, and R.L. and Y.S. carried out the epitaxial growth of $^{28}$Si. D.W. served as program manager. D.Sc., D.Si., T.K., V.A., and C.G. contributed to data interpretation and discussions. Y.C. synthesized the [TbPc$_2]^{-}$ crystal. M.R., K.D.G., and W.W. supervised and coordinated the project.

\clearpage
\appendix
\section{SiMOS qubit fabrication details}
\label{sec:fabrication_details}%
\begin{figure}[tb]
  \centering
  \includegraphics[width=0.5\columnwidth]{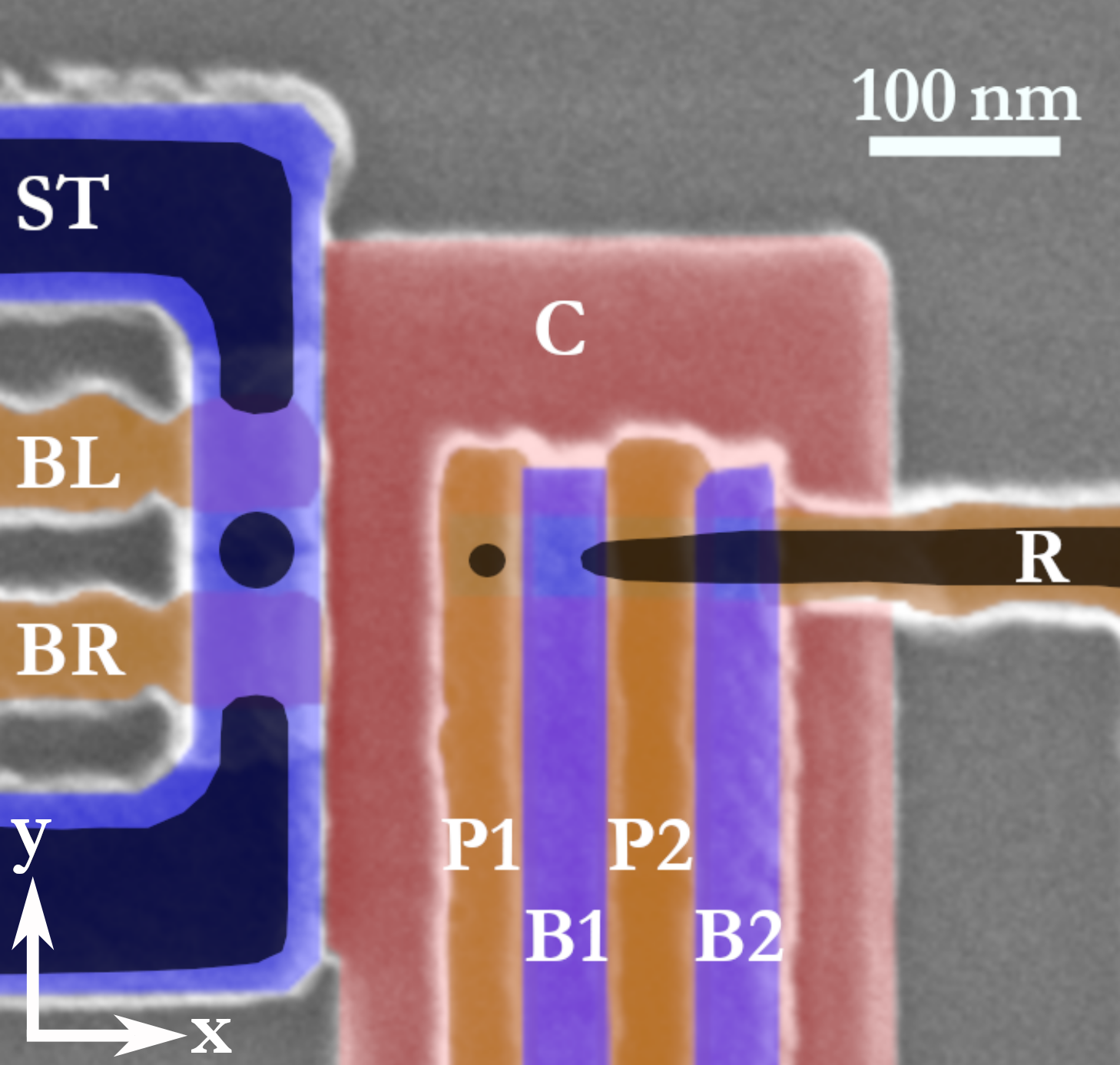}
  \caption{\textbf{Device architecture.}
    False-color scanning electron micrograph of the silicon quantum dot device prior to fabrication of the ESR antenna. The quantum dot, electron reservoir, and single electron transistor are formed in black shaded regions. The three poly-silicon gate layers are shown in blue (top), brown (middle), and red (bottom), respectively.
    }
  \label{fig:device}
\end{figure}
The SiMOS spin qubit device is fabricated at the $300\,\mathrm{mm}$ industrial pilot line of imec using advanced CMOS-compatible processing. The silicon wafer consists of a natural silicon substrate onto which a $200\,\mathrm{nm}$-thick isotopically enriched $^{28}\mathrm{Si}$ epilayer is deposited by chemical vapor deposition, yielding a residual $^{29}\mathrm{Si}$ concentration below $500\, \mathrm{ppm}$. A thermally grown $\mathrm{SiO}_2$ layer, $8\, \mathrm{nm}$ thick, serves as the gate oxide.

The gate architecture comprises three overlapping layers of doped poly-silicon, separated by $5\, \mathrm{nm}$ interlayer dielectric of sputtered $\mathrm{SiO}_2$. Patterning is achieved using electron beam lithography to define the gate electrodes. A microwave antenna made of titanium nitride is integrated above the gate stack and used to deliver electron spin resonance (ESR) excitations to the qubit \cite{Koppens2006}. A $250\, \mathrm{nm}$ $\mathrm{SiO}_2$ passivation layer encapsulates the gate stack, through which vias are etched to contact gate electrodes, ohmic leads, and the ESR antenna.

A false-colored scanning electron micrograph of a nominally identical device is shown in Fig.~\ref{fig:device}. The device layout supports the formation of a single-electron quantum dot under gate P1, tunnel-coupled to an electron reservoir formed beneath gates P2, B2 and R. An adjacent single-electron transistor (SET) serves as a charge sensor for spin readout. 
\section{Details of RPSS calibration and performance}
\label{sec:appendix_RPSS}
\paragraph*{Rabi frequency extraction.}
The power dependence of the resonance contrast in Fig.~\ref{fig:Methods}\textbf{b} was fitted using the Landau-Zener model:
\[
\mathrm{LZ}_\uparrow = 1 - \exp\left( -\frac{\pi^2 f_\mathrm{Rabi}^2}{|\frac{\partial}{\partial t}(\Delta f)|} \right),
\]
assuming $f_\mathrm{Rabi} \propto \sqrt{P}$, where \(P\) is the applied microwave power and $\Delta f$ the frequency detuning. The fit was limited to powers below 5\,dBm above which thermally activated tunneling depletes the dot prior to spin excitation. At 8\,dBm, the extracted Rabi frequency was \(f_\mathrm{Rabi} = 17.8 \pm 0.3\,\mathrm{kHz}\).
To verify adiabaticity at this power, we compared the chirp rate \(|\partial \Delta f / \partial t| = 0.2\,\mathrm{GHz/s}\) to \(f_\mathrm{Rabi}^2 = (17.8\,\mathrm{kHz})^2= 0.315\,\mathrm{GHz/s}\), confirming that the adiabatic condition ($|\partial \Delta f / \partial t| \ll f_\mathrm{Rabi}^2$) is satisfied above 8\,dBm which is indicated by the plateau in spin-up fraction in Fig.~\ref{fig:Methods}\textbf{b}.

\paragraph*{Sweep-rate limitations.}
The effect of chirp speed on signal contrast was investigated at constant drive power. Despite theoretical predictions of high inversion at chirp rates up to 1\,GHz/s, the measured visibility dropped rapidly due to finite tunneling rates and nonuniform time-frequency binning. Each frequency bin maps to a shorter time interval at higher sweep rates, lowering the probability of capturing tunneling events within each bin. Therefore, slower sweeps are preferred for high-fidelity detection, albeit at the cost of longer measurement times.

\begin{figure}[tb]
  \centering
  \includegraphics[width=\columnwidth]{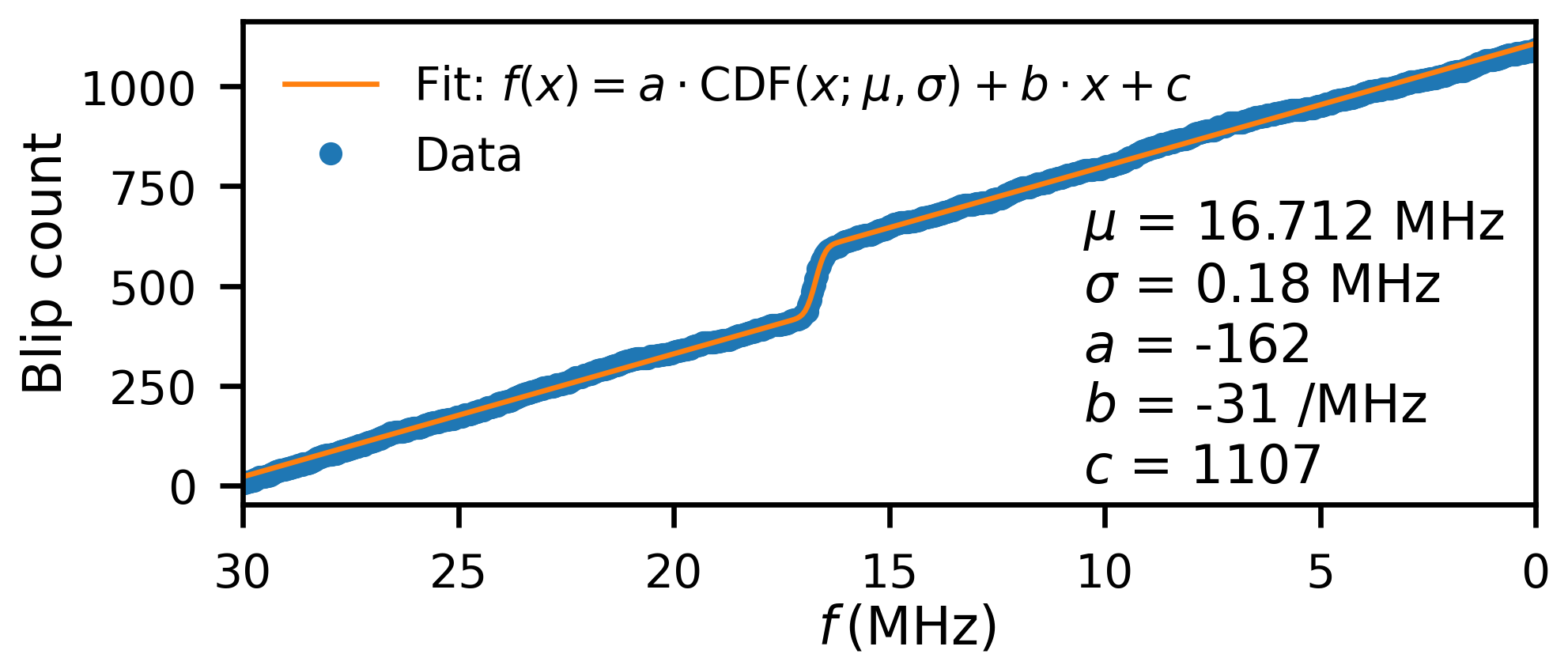}
  \caption{\textbf{Cumulative blip count analysis.} 
    The same dataset shown in Fig.~\ref{fig:Methods}\textbf{d} is here analyzed without the 1\,MHz binning. All blip events are cumulatively summed as a function of time and then mapped to frequency. The data are fitted using a cumulative distribution function (CDF) combined with a linear background to account for thermal blip events. The fit yields a resonance frequency of $f = 16.7$\,MHz offset by 18.57\,GHz, with a standard deviation $\sigma = 0.2$\,MHz, providing finer frequency resolution than the 1\,MHz-binned method.}
  \label{fig:cum_count}
\end{figure}

\paragraph*{Alternative analysis methods.}
We explored extracting the resonance position by fitting the cumulative distribution of blip times, rather than histogramming them to locate the point of steepest slope, corresponding to the RAP transition (Fig.~\ref{fig:cum_count}). While the cumulative fit yielded a nominally finer frequency uncertainty of $0.18\,\mathrm{MHz}$, this method proved less robust in practice due to variable background rates and temporal jitter in blip onset. For this reason, binned histograms were adopted throughout the analysis.

\section{[TbPc$_2]^{-}$ crystal placement}
\label{subsec:TbPc2}
\begin{figure}[tb]
    \centering
    \includegraphics[width=0.6\columnwidth]{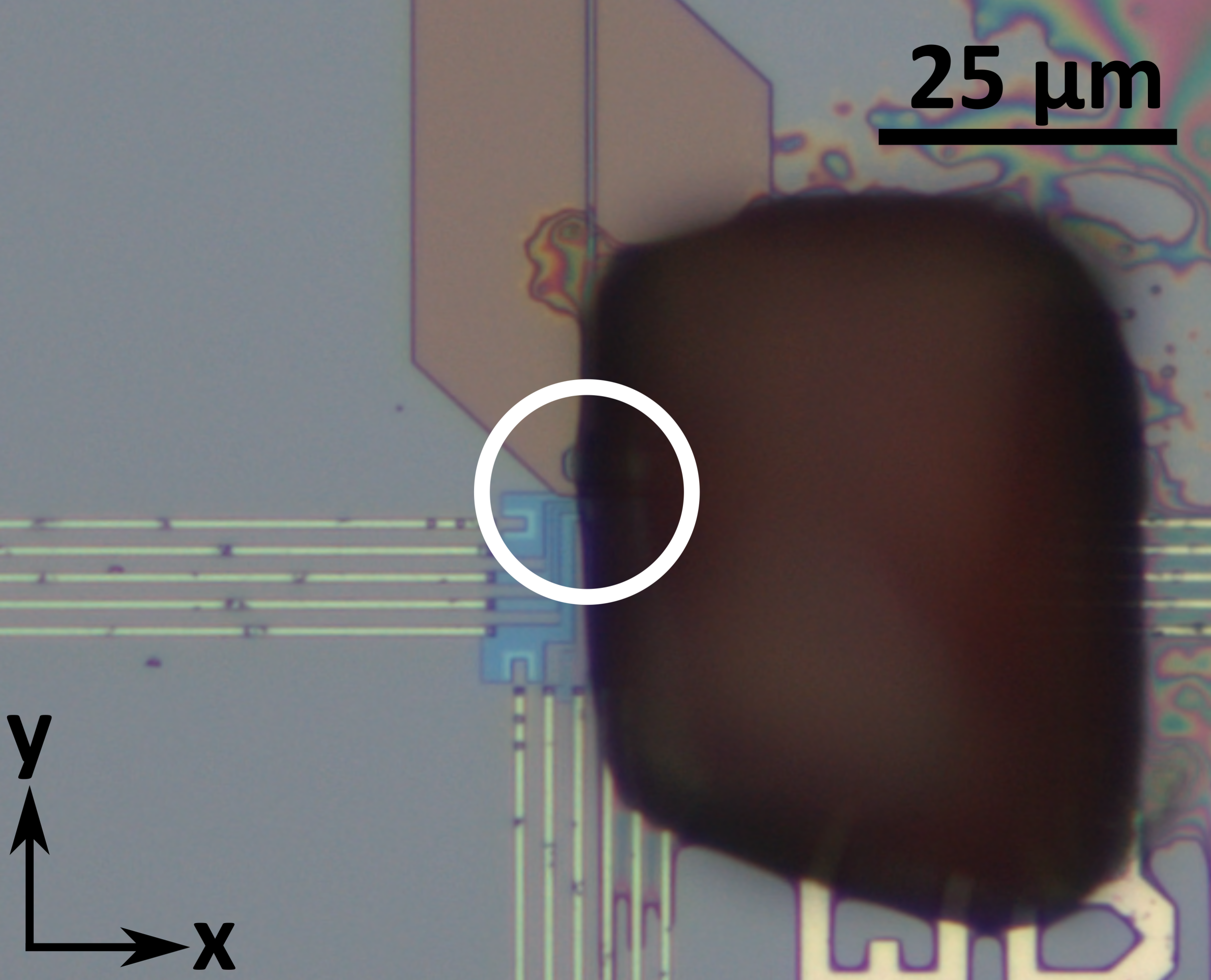}
    \caption{\textbf{Crystal placement.}
    Optical image of a diluted [TbPc$_2]^{-}$ ensemble placed on top of the device. The spin qubit is located at the center of the white circle. The microwave antenna enters from the top, while DC gate lines connect from the other sides.}
  \label{fig:Crystal_placement}
\end{figure}
Terbium bis(phthalocyaninato) (TbPc\(_2\)) exists in two charge states relevant for quantum experiments: the neutral radical form and the anionic form [TbPc\(_2\)]\(^-\). The neutral species contains a ligand-centered unpaired electron, which can mediate long-range exchange coupling between molecules. While this radical can be exploited for certain spin readout schemes at the single-molecule level via the spin cascade effect~\cite{Thiele2013, Thiele2014, Godfrin2017}, it also introduces additional coupling pathways that complicate the study of the intrinsic properties of the Tb\(^{3+}\) electronic and nuclear spin states. In the anionic form, the ligand radical is absent, eliminating radical-mediated coupling and thereby allowing the magnetization dynamics of the Tb\(^{3+}\) ion to be probed in isolation. For this reason, the experiments presented here employ the anionic form [TbPc\(_2\)]\(^-\), which is more suitable for ensemble measurements of intrinsic molecular spin behavior.

To generate a detectable magnetic stray field at the quantum dot site, a microcrystal containing an ensemble of [TbPc$_2]^{-}$ molecules is used. The collective dipolar field of many Tb\(^{3+}\) ions enhances the overall magnetic signal, which would be too weak to resolve for a single SMM at the current qubit-surface distance. To preserve the single-molecule magnetic character and minimize intermolecular interactions, the crystal is grown in a diluted form with 5\% [TbPc$_2]^{-}$ and 95\% diamagnetic YPc\(_2\) units.

\begin{figure}[tb]
  \centering
  \includegraphics[width=\columnwidth]{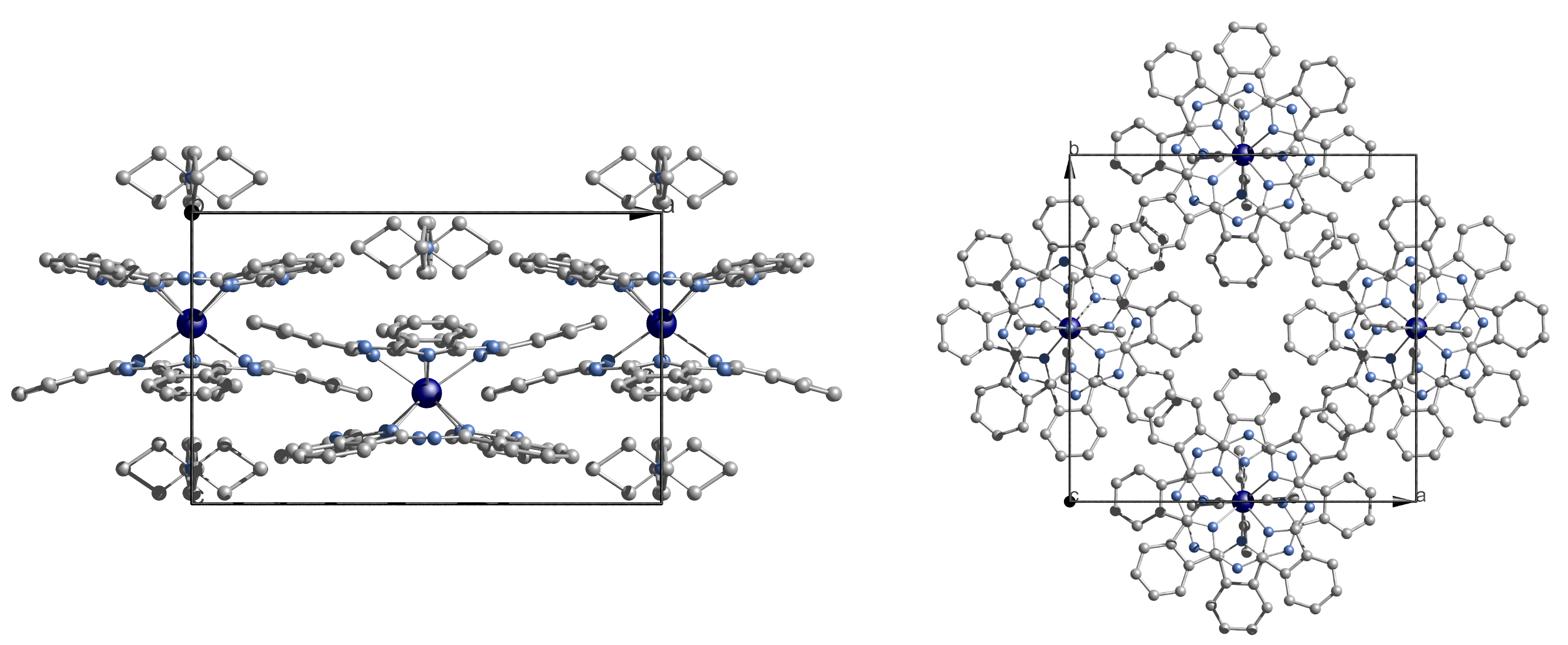}
  \caption{\textbf{Crystal packing of diluted [TbPc\(_2\)]\(^-\) in the unit cell.} 
  Molecular arrangement within a unit cell, with additional Tetraethylammonium (TEA) ions incorporated to maintain charge neutrality viewed from the side (\textbf{left}) and along the stacking axis (\textbf{right}). Color code: Tb/Y (dark blue), N (light blue), C (gray), H omitted.}
  \label{fig:TbPc2_coordination}
\end{figure}

A single [TbPc$_2]^{-}$ crystal, typically $50~\upmu$m in size, is placed on the device surface after fabrication using a micro-manipulator under ambient conditions. The orientation of the crystal’s magnetic easy axis can be readily adjusted, as it coincides with one of the crystal faces. As shown in Fig.~\ref{fig:TbPc2_coordination}, the molecular symmetry axis is parallel to a unit cell vector, providing a clear crystallographic reference for alignment. In the experiments, the easy axis of magnetization is oriented perpendicular to the chip plane (along the $z$-axis), as illustrated in Fig.~\ref{fig:Schematic_coupling}. A static external magnetic field, used to Zeeman-split the spin qubit levels, is applied in-plane (along the $x$-axis), thus perpendicular to the [TbPc$_2]^{-}$ easy axis. This configuration aligns the field along the magnetic hard axis, minimizing its influence on the SMM’s magnetic switching behavior.

To optimize the qubit’s sensitivity to magnetic switching of the [TbPc$_2]^{-}$, the crystal is placed slightly offset along the $x$-direction such that only the edge overlaps the quantum dot region. This lateral displacement enhances the $x$-component of the dipolar field experienced by the quantum dot, thereby increasing the effective Zeeman shift of the spin qubit during magnetic switching events.

Nuclear spin transitions within the molecule remained unresolved. We attribute this to inhomogeneous broadening arising from a distribution of easy-axis orientations within the crystal. In the presence of a strong transverse field of 0.66\,T, small angular variations translate into different resonance conditions across the ensemble, smearing the hyperfine-resolved features. Distributions of easy-axis tilt angles in TbPc$_2$ ensembles have been directly resolved by torque magnetometry in thin films~\cite{Perfetti2016}. Moreover, variations in molecular packing and local environment are known to modify the magnetic anisotropy and tunneling behavior of TbPc$_2$~\cite{Yamabayashi2017,Malavolti2013}, consistent with the broadening observed here.

\section{3D vector magnet calibration}
\label{subsec:magnet_calib}
A custom-built 3D vector magnet, constructed from NbTi superconducting wire is used to control the magnetic field vector at the location of the device. For the present experiment, the objective is to vary the direction of the applied magnetic field while maintaining a constant total field amplitude. This ensures that the qubit resonance frequency remains fixed during angular sweeps.

The magnetic field is tuned freely within the $yz$-plane up to a field magnitude of 0.28\,T. A compensation field is simultaneously applied along the $x$-axis to maintain a constant total magnetic field of approximately 0.66\,T. The angle $\theta$ defines the polar angle of the magnetic field vector within the $yz$-plane, defined in spherical coordinates.

To account for coil misalignments relative to the Cartesian coordinate system, correction angles are introduced: $\Delta_\mathrm{xy} = 0.75^\circ$ for the $x$-$y$ coil plane and $\Delta_\mathrm{xz} = -2.6^\circ$ for the $x$-$z$ plane. These values quantify the deviation from orthogonality.  For simplicity, the $y$-$z$ coil configuration is assumed to be exactly orthogonal, as residual misalignments in this plane are less critical given the dominant contribution of the $x$-axis field (which exceeds the $yz$ components by more than a factor of 2 in the sweeps shown.

The coil calibration parameters, comprised of current-to-field conversion factors and alignment corrections, are determined using a feedback protocol in which the microwave drive frequency is dynamically adjusted after each measurement to keep the qubit resonance centered within the 100\,MHz detection window.

Assuming a total magnetic field magnitude \( A_\mathrm{const} \), and a projected in-plane component \( A_\mathrm{pol} \) at angle \( \theta \) in the $yz$-plane, the field components of the vector magnet along each of its axis are given by: 
\begin{align*} 
A_x &= \sqrt{A_\mathrm{const}^2 - A_\mathrm{pol}^2} 
+ A_\mathrm{pol} \left[\cos(\theta) \cdot \tan(\Delta_\mathrm{xz}) \right. \\
&\quad \left. + \sin(\theta) \cdot \tan(\Delta_\mathrm{xy})
\right] \\
A_y &= A_\mathrm{pol} \cdot \sin(\theta) \cdot \cos(\Delta_\mathrm{xy})^{-1} \\
A_z &= A_\mathrm{pol} \cdot \cos(\theta) \cdot \cos(\Delta_\mathrm{xz})^{-1} \,.
\end{align*}

\section{Setup}
\label{subsec:setup}
All measurements are performed in a Qinu Version L dilution refrigerator operating at a base temperature below $48\, \mathrm{mK}$. The effective electron temperature is determined to be $\leq 130\, \mathrm{mK}$, extracted via the thermal broadening of the
0→1 qubit charge transition as a function of mixing chamber temperature \cite{Beenakker1991, Maradan2014}.

DC lines are filtered by two-stage low-pass RC filters anchored at the mixing chamber stage, with cutoff frequencies of $250\, \mathrm{Hz}$ for static gate voltages and $40\, \mathrm{kHz}$ for sensor current lines. Each line additionally includes a Mini-Circuits LFCN-80+ low-pass filter to suppress intermediate-frequency noise between $225\, \mathrm{MHz}$ and $4.5\, \mathrm{GHz}$. A resistive bias-tee is employed on the plunger gate (P1).

To improve voltage resolution on the DC gates, the plunger voltage is delivered via a high-frequency line, which includes $33\, \mathrm{dB}$ attenuation at room temperature. The signal is capacitively coupled through a resistive bias-tee. To compensate for discharge effects in the bias-tee’s series capacitor, a linear voltage ramp is superimposed onto the pulse. Pulsed operation of the plunger becomes unnecessary if higher-resolution DC control is available.

Microwave signals are delivered with a room-temperature attenuation of $61 \, \mathrm{dB}$ at $18.5\,\mathrm{GHz}$ consisting of the attenuation of cupronickel coaxial cables, inline attenuation, and mixer loss. Static gate voltages are supplied by an ADwin Pro II system (Jaeger Messtechnik), while charge sensor current is converted to voltage at room temperature using a Basel Precision Instruments SP983c. The resulting voltage is digitized by the Pro II-AIn-F-4/16 analog input module of the ADwin system.

Frequency-chirped microwave pulses for RAP excitation are synthesized using a Zurich Instruments HDAWG, upconverted by mixing with a Keysight N5183B microwave source and a Marki Microwave SSB-0618 single-sideband mixer.

\section{Control Measurements without SMM}
\label{sec:hysteresis_no_SMM}
\begin{figure}[tb]
  \centering
  \includegraphics[width=\columnwidth]{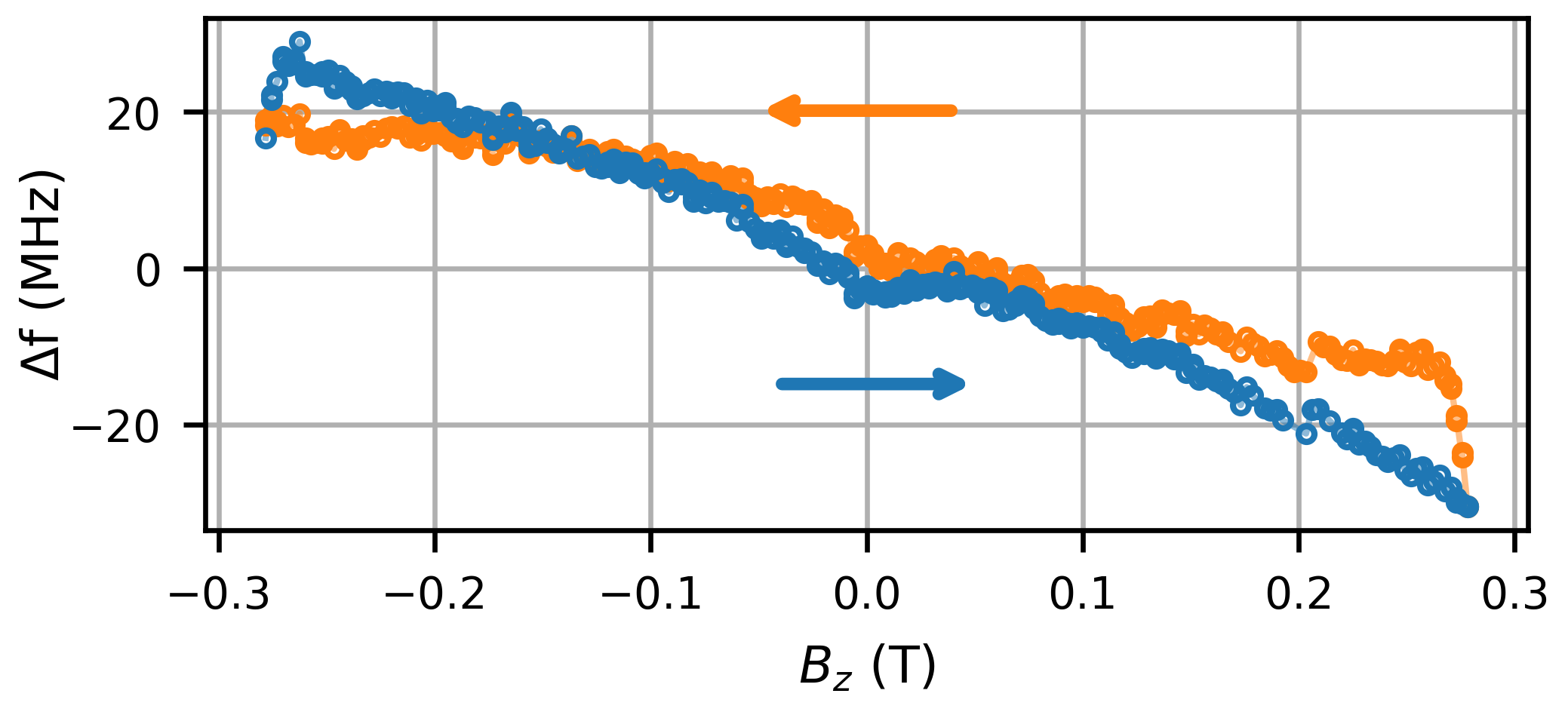}
  \caption{\textbf{Hysteresis measurement in absence of the [TbPc$_2]^{-}$ crystal using the RPSS protocol.}
    Qubit resonance frequency (offset by 18.435\,GHz) as a function of the magnetic field component along the $z$-axis, recorded during angular sweeps of the external field in the $xz$-plane at 48\,mK. Arrows indicate the sweep direction. A residual tilt in the trace arises from imperfect calibration of the vector field compensation. Coil-induced artifacts remain visible at the sweep reversal points $\pm$0.28\,T.
    }
  \label{fig:hysteresis_no_SMM}
\end{figure}

To isolate magnetic signals originating from the [TbPc$_2]^{-}$ crystal, we performed control measurements with the molecular sample removed. Figure~\ref{fig:hysteresis_no_SMM} shows the qubit resonance frequency during angular magnetic field sweeps in the $xz$-plane at 48\,mK. A residual tilt in the trace is attributed to imperfect vector field compensation. The pronounced 30\,MHz loop observed in the presence of the [TbPc$_2]^{-}$ crystal (Fig.~\ref{fig:Results}\textbf{a}) is no longer visible, confirming that the dominant magnetic response originates from the on-chip sample.

The residual anomalies at the sweep turning points are attributed to flux vortex rearrangements within the type-II superconducting NbTi vector magnet. These artifacts consistently appear upon reversal of the sweep direction or when the magnetic field crosses zero, and they persist irrespective of temperature or sample presence. Each field overshoot gradually relaxes with unidirectional ramping or repeated cycling around the target field value.

This underscores the need to account for instrumental background effects in cryogenic magnetometry experiments involving superconducting coil systems.

\section{T1 and T2* of the SiMOS qubit}
\label{sec:T1_T2}
\begin{figure}[tb]
  \centering
  \includegraphics[width=\columnwidth]{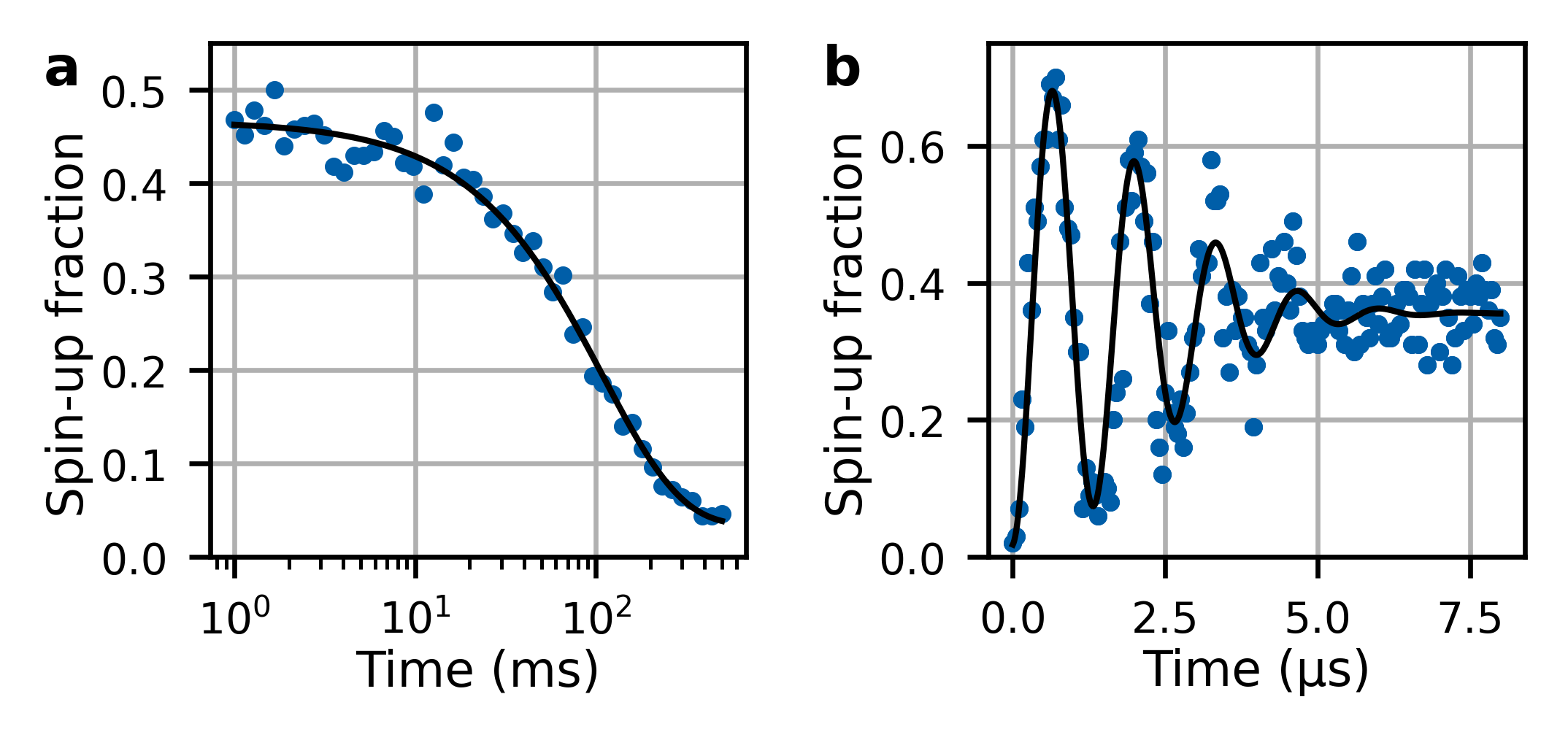}
  \caption{\textbf{Qubit relaxation and coherence at 0.66\,T in absence of a SMM.}
    \textbf{a} Spin relaxation time $T_1$ measured using the Elzerman pulse sequence \cite{Elzerman2004}. Each data point (blue) represents an average over 100 single-shot measurements. An exponential fit (black) yields $T_1 = (110 \pm 7)$\,ms.
    \textbf{b} Ramsey interference measurement performed with a detuning of 739\,kHz.   The fitted decay (black) gives a coherence time of $T_2^* = (3.1 \pm 0.1)\,\upmu$s, with 100 averages per point.
    }
  \label{fig:T1_T2}
\end{figure}

Using the Elzerman pulse scheme~\cite{Elzerman2004}, the spin relaxation time $T_1$ of the silicon qubit was measured to be $110 \pm 7$\,ms (Fig.~\ref{fig:T1_T2}\textbf{a}) at a field value of $0.66\, \mathrm{T}$. The Ramsey decoherence time $T_2^*$ was extracted as $3.1 \pm 0.1\,\upmu$s from a Ramsey interference experiment acquired over 21~min (Fig.~\ref{fig:T1_T2}\textbf{b}). During the Ramsey sequence, the microwave pulses were applied while both spin states remained below the Fermi level of the reservoir.

\section{Pulse schemes for sensing}
\label{sec:sim_pulses}%
\begin{figure}[tb]
  \centering
  \includegraphics[width=\columnwidth]{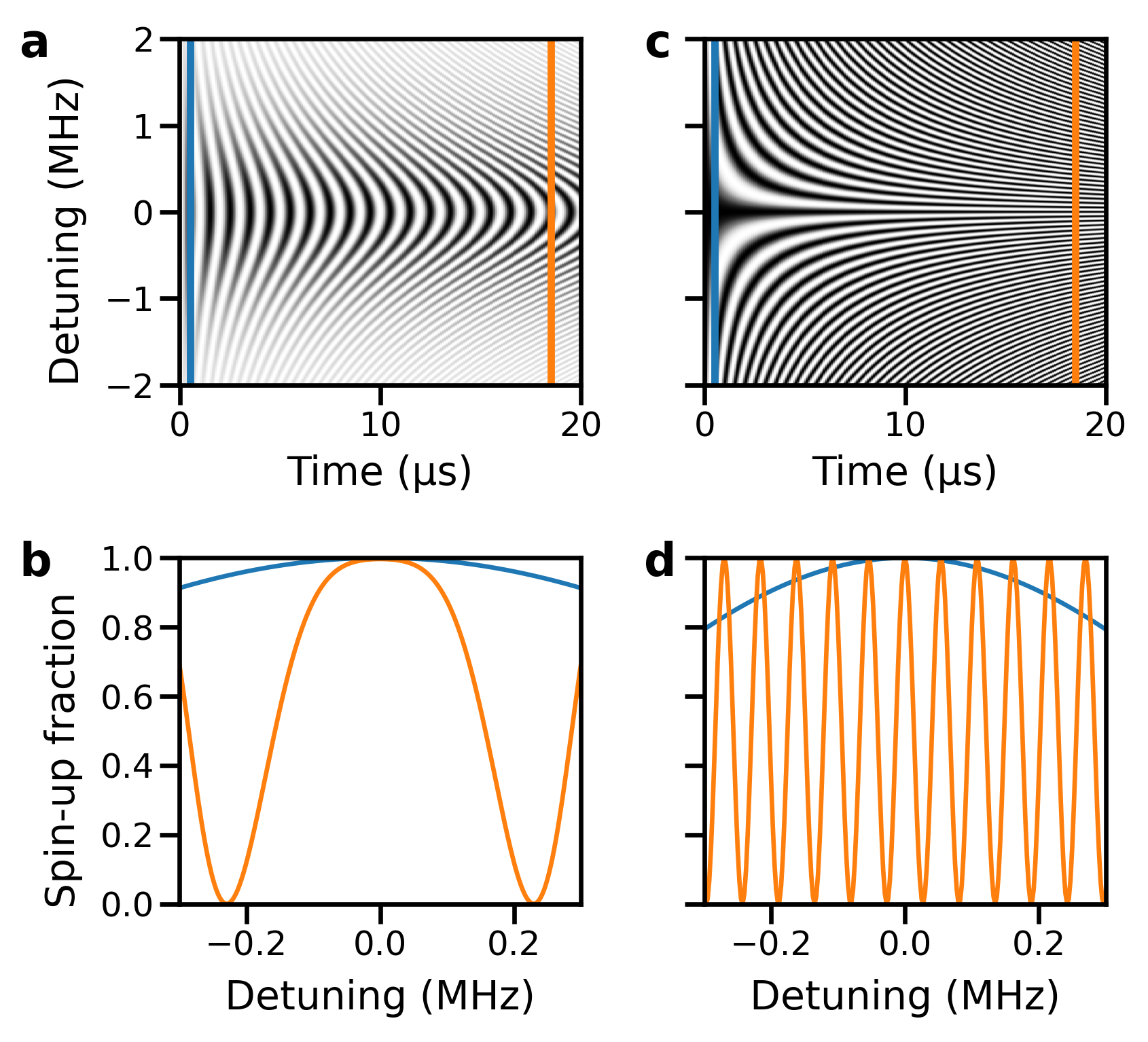}
    \caption{\textbf{Sensitivity of pulse protocols for Larmor frequency determination.} 
    \textbf{a, c} Time-resolved spin dynamics under a Rabi pulse (\textbf{a}) and a Ramsey sequence (\textbf{c}), simulated with a Rabi frequency of 1\,MHz in an ideal, decoherence-free system. The spin-up fraction is represented in grayscale (white = 0\%, black = 100\%). Vertical lines indicate two representative readout times. 
    \textbf{b, d} Spin-up probability as a function of detuning, extracted at the readout times from \textbf{a} and \textbf{c}. The Ramsey sequence (\textbf{d}) shows sharper and more densely spaced oscillations than the Rabi protocol (\textbf{b}), reflecting its higher sensitivity to resonance shifts. Longer pulse durations (orange) yield greater frequency resolution than shorter pulses (blue), highlighting the trade-off between spectral precision and measurement time in qubit-based magnetometry.}
  \label{fig:sim_pulses}
\end{figure}

The choice of pulse sequence directly determines the achievable frequency resolution and magnetic sensitivity. To evaluate the advantages of more advanced protocols over simple continuous driving, we numerically compare Rabi and Ramsey pulse schemes for determining the qubit's Larmor frequency.

Figure~\ref{fig:sim_pulses} presents simulated spin dynamics under both protocols. As shown in Fig.~\ref{fig:sim_pulses}\textbf{d}, the Ramsey signal exhibits higher fringe density and greater sensitivity to small detunings compared to the broader and lower-contrast response of the Rabi protocol in Fig.~\ref{fig:sim_pulses}\textbf{b}.

The simulations also illustrate the trade-off between frequency resolution and measurement time: longer pulse durations improve spectral selectivity but reduce temporal resolution. For magnetometry applications targeting narrow-band or weak-field signals-such as those expected from single-molecule magnets- dynamical decoupling sequences offer a clear advantage over simpler protocols. These techniques will be central to future implementations of hybrid qubit-molecule sensors aimed at resolving individual spin transitions with high precision.

\clearpage
\bibliography{references} 

\begin{thebibliography}{43}%
\makeatletter
\providecommand \@ifxundefined [1]{%
 \@ifx{#1\undefined}
}%
\providecommand \@ifnum [1]{%
 \ifnum #1\expandafter \@firstoftwo
 \else \expandafter \@secondoftwo
 \fi
}%
\providecommand \@ifx [1]{%
 \ifx #1\expandafter \@firstoftwo
 \else \expandafter \@secondoftwo
 \fi
}%
\providecommand \natexlab [1]{#1}%
\providecommand \enquote  [1]{``#1''}%
\providecommand \bibnamefont  [1]{#1}%
\providecommand \bibfnamefont [1]{#1}%
\providecommand \citenamefont [1]{#1}%
\providecommand \href@noop [0]{\@secondoftwo}%
\providecommand \href [0]{\begingroup \@sanitize@url \@href}%
\providecommand \@href[1]{\@@startlink{#1}\@@href}%
\providecommand \@@href[1]{\endgroup#1\@@endlink}%
\providecommand \@sanitize@url [0]{\catcode `\\12\catcode `\$12\catcode `\&12\catcode `\#12\catcode `\^12\catcode `\_12\catcode `\%12\relax}%
\providecommand \@@startlink[1]{}%
\providecommand \@@endlink[0]{}%
\providecommand \url  [0]{\begingroup\@sanitize@url \@url }%
\providecommand \@url [1]{\endgroup\@href {#1}{\urlprefix }}%
\providecommand \urlprefix  [0]{URL }%
\providecommand \Eprint [0]{\href }%
\providecommand \doibase [0]{https://doi.org/}%
\providecommand \selectlanguage [0]{\@gobble}%
\providecommand \bibinfo  [0]{\@secondoftwo}%
\providecommand \bibfield  [0]{\@secondoftwo}%
\providecommand \translation [1]{[#1]}%
\providecommand \BibitemOpen [0]{}%
\providecommand \bibitemStop [0]{}%
\providecommand \bibitemNoStop [0]{.\EOS\space}%
\providecommand \EOS [0]{\spacefactor3000\relax}%
\providecommand \BibitemShut  [1]{\csname bibitem#1\endcsname}%
\let\auto@bib@innerbib\@empty
%</preamble>
\bibitem [{\citenamefont {Atzori}\ and\ \citenamefont {Sessoli}(2019)}]{Atzori2019}%
  \BibitemOpen
  \bibfield  {author} {\bibinfo {author} {\bibfnamefont {M.}~\bibnamefont {Atzori}}\ and\ \bibinfo {author} {\bibfnamefont {R.}~\bibnamefont {Sessoli}},\ }\bibfield  {title} {\bibinfo {title} {The second quantum revolution: Role and challenges of molecular chemistry},\ }\href {https://doi.org/10.1021/jacs.9b00984} {\bibfield  {journal} {\bibinfo  {journal} {Journal of the American Chemical Society}\ }\textbf {\bibinfo {volume} {141}},\ \bibinfo {pages} {11339} (\bibinfo {year} {2019})}\BibitemShut {NoStop}%
\bibitem [{\citenamefont {Wernsdorfer}\ and\ \citenamefont {Ruben}(2019)}]{Wernsdorfer2019}%
  \BibitemOpen
  \bibfield  {author} {\bibinfo {author} {\bibfnamefont {W.}~\bibnamefont {Wernsdorfer}}\ and\ \bibinfo {author} {\bibfnamefont {M.}~\bibnamefont {Ruben}},\ }\bibfield  {title} {\bibinfo {title} {Synthetic hilbert space engineering of molecular qudits: Isotopologue chemistry},\ }\href {https://doi.org/10.1002/adma.201806687} {\bibfield  {journal} {\bibinfo  {journal} {Advanced Materials}\ }\textbf {\bibinfo {volume} {31}},\ \bibinfo {pages} {1806687} (\bibinfo {year} {2019})}\BibitemShut {NoStop}%
\bibitem [{\citenamefont {Moreno‑Pineda}\ and\ \citenamefont {Wernsdorfer}(2024)}]{MorenoPineda2024}%
  \BibitemOpen
  \bibfield  {author} {\bibinfo {author} {\bibfnamefont {E.}~\bibnamefont {Moreno‑Pineda}}\ and\ \bibinfo {author} {\bibfnamefont {W.}~\bibnamefont {Wernsdorfer}},\ }\bibfield  {title} {\bibinfo {title} {Magnetic molecules as building blocks for quantum technologies},\ }\href {https://doi.org/10.1002/qute.202300367} {\bibfield  {journal} {\bibinfo  {journal} {Advanced Quantum Technologies}\ }\textbf {\bibinfo {volume} {7}},\ \bibinfo {pages} {2300367} (\bibinfo {year} {2024})}\BibitemShut {NoStop}%
\bibitem [{\citenamefont {Moreno‑Pineda}\ \emph {et~al.}(2018)\citenamefont {Moreno‑Pineda}, \citenamefont {Godfrin}, \citenamefont {Balestro}, \citenamefont {Wernsdorfer},\ and\ \citenamefont {Ruben}}]{MorenoPineda2018}%
  \BibitemOpen
  \bibfield  {author} {\bibinfo {author} {\bibfnamefont {E.}~\bibnamefont {Moreno‑Pineda}}, \bibinfo {author} {\bibfnamefont {C.}~\bibnamefont {Godfrin}}, \bibinfo {author} {\bibfnamefont {F.}~\bibnamefont {Balestro}}, \bibinfo {author} {\bibfnamefont {W.}~\bibnamefont {Wernsdorfer}},\ and\ \bibinfo {author} {\bibfnamefont {M.}~\bibnamefont {Ruben}},\ }\bibfield  {title} {\bibinfo {title} {Molecular spin qudits for quantum algorithms},\ }\href {https://doi.org/10.1039/C5CS00933B} {\bibfield  {journal} {\bibinfo  {journal} {Chemical Society Reviews}\ }\textbf {\bibinfo {volume} {47}},\ \bibinfo {pages} {501} (\bibinfo {year} {2018})}\BibitemShut {NoStop}%
\bibitem [{\citenamefont {Janković}\ \emph {et~al.}(2024)\citenamefont {Janković}, \citenamefont {Hartmann}, \citenamefont {Ruben},\ and\ \citenamefont {Hervieux}}]{Jankovic2024}%
  \BibitemOpen
  \bibfield  {author} {\bibinfo {author} {\bibfnamefont {D.}~\bibnamefont {Janković}}, \bibinfo {author} {\bibfnamefont {J.}~\bibnamefont {Hartmann}}, \bibinfo {author} {\bibfnamefont {M.}~\bibnamefont {Ruben}},\ and\ \bibinfo {author} {\bibfnamefont {P.}~\bibnamefont {Hervieux}},\ }\bibfield  {title} {\bibinfo {title} {Noisy qudit vs multiple qubits: Conditions on gate efficiency for enhancing fidelity},\ }\href {https://doi.org/10.1038/s41534-024-00829-6} {\bibfield  {journal} {\bibinfo  {journal} {npj Quantum Information}\ }\textbf {\bibinfo {volume} {10}},\ \bibinfo {pages} {59} (\bibinfo {year} {2024})}\BibitemShut {NoStop}%
\bibitem [{\citenamefont {Ishikawa}\ \emph {et~al.}(2003)\citenamefont {Ishikawa}, \citenamefont {Sugita}, \citenamefont {Ishikawa}, \citenamefont {Koshihara},\ and\ \citenamefont {Kaizu}}]{Ishikawa2003}%
  \BibitemOpen
  \bibfield  {author} {\bibinfo {author} {\bibfnamefont {N.}~\bibnamefont {Ishikawa}}, \bibinfo {author} {\bibfnamefont {M.}~\bibnamefont {Sugita}}, \bibinfo {author} {\bibfnamefont {T.}~\bibnamefont {Ishikawa}}, \bibinfo {author} {\bibfnamefont {S.~Y.}\ \bibnamefont {Koshihara}},\ and\ \bibinfo {author} {\bibfnamefont {Y.}~\bibnamefont {Kaizu}},\ }\bibfield  {title} {\bibinfo {title} {Lanthanide double-decker complexes functioning as magnets at the single-molecular level},\ }\href {https://doi.org/10.1021/ja029629n} {\bibfield  {journal} {\bibinfo  {journal} {Journal of the American Chemical Society}\ }\textbf {\bibinfo {volume} {125}},\ \bibinfo {pages} {8694} (\bibinfo {year} {2003})}\BibitemShut {NoStop}%
\bibitem [{\citenamefont {Thiele}\ \emph {et~al.}(2013)\citenamefont {Thiele}, \citenamefont {Vincent}, \citenamefont {Holzmann}, \citenamefont {Klyatskaya}, \citenamefont {Ruben}, \citenamefont {Balestro},\ and\ \citenamefont {Wernsdorfer}}]{Thiele2013}%
  \BibitemOpen
  \bibfield  {author} {\bibinfo {author} {\bibfnamefont {S.}~\bibnamefont {Thiele}}, \bibinfo {author} {\bibfnamefont {R.}~\bibnamefont {Vincent}}, \bibinfo {author} {\bibfnamefont {M.}~\bibnamefont {Holzmann}}, \bibinfo {author} {\bibfnamefont {S.}~\bibnamefont {Klyatskaya}}, \bibinfo {author} {\bibfnamefont {M.}~\bibnamefont {Ruben}}, \bibinfo {author} {\bibfnamefont {F.}~\bibnamefont {Balestro}},\ and\ \bibinfo {author} {\bibfnamefont {W.}~\bibnamefont {Wernsdorfer}},\ }\bibfield  {title} {\bibinfo {title} {Electrical readout of individual nuclear spin trajectories in a single-molecule magnet spin transistor},\ }\href {https://doi.org/10.1103/PhysRevLett.111.037203} {\bibfield  {journal} {\bibinfo  {journal} {Physical Review Letters}\ }\textbf {\bibinfo {volume} {111}},\ \bibinfo {pages} {037203} (\bibinfo {year} {2013})}\BibitemShut {NoStop}%
\bibitem [{\citenamefont {Thiele}\ \emph {et~al.}(2014)\citenamefont {Thiele}, \citenamefont {Balestro}, \citenamefont {Ballou}, \citenamefont {Klyatskaya}, \citenamefont {Ruben},\ and\ \citenamefont {Wernsdorfer}}]{Thiele2014}%
  \BibitemOpen
  \bibfield  {author} {\bibinfo {author} {\bibfnamefont {S.}~\bibnamefont {Thiele}}, \bibinfo {author} {\bibfnamefont {F.}~\bibnamefont {Balestro}}, \bibinfo {author} {\bibfnamefont {R.}~\bibnamefont {Ballou}}, \bibinfo {author} {\bibfnamefont {S.}~\bibnamefont {Klyatskaya}}, \bibinfo {author} {\bibfnamefont {M.}~\bibnamefont {Ruben}},\ and\ \bibinfo {author} {\bibfnamefont {W.}~\bibnamefont {Wernsdorfer}},\ }\bibfield  {title} {\bibinfo {title} {Electrically driven nuclear spin resonance in single-molecule magnets},\ }\href {https://doi.org/10.1126/science.1249802} {\bibfield  {journal} {\bibinfo  {journal} {Science}\ }\textbf {\bibinfo {volume} {344}},\ \bibinfo {pages} {1135} (\bibinfo {year} {2014})}\BibitemShut {NoStop}%
\bibitem [{\citenamefont {Godfrin}\ \emph {et~al.}(2017)\citenamefont {Godfrin}, \citenamefont {Ferhat}, \citenamefont {Ballou}, \citenamefont {Klyatskaya}, \citenamefont {Ruben}, \citenamefont {Wernsdorfer},\ and\ \citenamefont {Balestro}}]{Godfrin2017}%
  \BibitemOpen
  \bibfield  {author} {\bibinfo {author} {\bibfnamefont {C.}~\bibnamefont {Godfrin}}, \bibinfo {author} {\bibfnamefont {A.}~\bibnamefont {Ferhat}}, \bibinfo {author} {\bibfnamefont {R.}~\bibnamefont {Ballou}}, \bibinfo {author} {\bibfnamefont {S.}~\bibnamefont {Klyatskaya}}, \bibinfo {author} {\bibfnamefont {M.}~\bibnamefont {Ruben}}, \bibinfo {author} {\bibfnamefont {W.}~\bibnamefont {Wernsdorfer}},\ and\ \bibinfo {author} {\bibfnamefont {F.}~\bibnamefont {Balestro}},\ }\bibfield  {title} {\bibinfo {title} {Operating quantum states in single magnetic molecules: Implementation of grover's quantum algorithm},\ }\href {https://doi.org/10.1103/PhysRevLett.119.187702} {\bibfield  {journal} {\bibinfo  {journal} {Physical Review Letters}\ }\textbf {\bibinfo {volume} {119}},\ \bibinfo {pages} {187702} (\bibinfo {year} {2017})}\BibitemShut {NoStop}%
\bibitem [{\citenamefont {Godfrin}\ \emph {et~al.}(2018)\citenamefont {Godfrin}, \citenamefont {Ballou}, \citenamefont {Bonet}, \citenamefont {Ruben}, \citenamefont {Klyatskaya}, \citenamefont {Wernsdorfer},\ and\ \citenamefont {Balestro}}]{Godfrin2018}%
  \BibitemOpen
  \bibfield  {author} {\bibinfo {author} {\bibfnamefont {C.}~\bibnamefont {Godfrin}}, \bibinfo {author} {\bibfnamefont {R.}~\bibnamefont {Ballou}}, \bibinfo {author} {\bibfnamefont {E.}~\bibnamefont {Bonet}}, \bibinfo {author} {\bibfnamefont {M.}~\bibnamefont {Ruben}}, \bibinfo {author} {\bibfnamefont {S.}~\bibnamefont {Klyatskaya}}, \bibinfo {author} {\bibfnamefont {W.}~\bibnamefont {Wernsdorfer}},\ and\ \bibinfo {author} {\bibfnamefont {F.}~\bibnamefont {Balestro}},\ }\bibfield  {title} {\bibinfo {title} {Generalized ramsey interferometry explored with a single nuclear spin qudit},\ }\href {https://doi.org/10.1038/s41534-018-0101-3} {\bibfield  {journal} {\bibinfo  {journal} {npj Quantum Information}\ }\textbf {\bibinfo {volume} {4}},\ \bibinfo {pages} {53} (\bibinfo {year} {2018})}\BibitemShut {NoStop}%
\bibitem [{\citenamefont {Urdampilleta}\ \emph {et~al.}(2011)\citenamefont {Urdampilleta}, \citenamefont {Klyatskaya}, \citenamefont {Cleuziou}, \citenamefont {Ruben},\ and\ \citenamefont {Wernsdorfer}}]{Urdampilleta2011}%
  \BibitemOpen
  \bibfield  {author} {\bibinfo {author} {\bibfnamefont {M.}~\bibnamefont {Urdampilleta}}, \bibinfo {author} {\bibfnamefont {S.}~\bibnamefont {Klyatskaya}}, \bibinfo {author} {\bibfnamefont {J.}~\bibnamefont {Cleuziou}}, \bibinfo {author} {\bibfnamefont {M.}~\bibnamefont {Ruben}},\ and\ \bibinfo {author} {\bibfnamefont {W.}~\bibnamefont {Wernsdorfer}},\ }\bibfield  {title} {\bibinfo {title} {Supramolecular spin valves},\ }\href {https://doi.org/10.1038/nmat3050} {\bibfield  {journal} {\bibinfo  {journal} {Nature Materials}\ }\textbf {\bibinfo {volume} {10}},\ \bibinfo {pages} {502} (\bibinfo {year} {2011})}\BibitemShut {NoStop}%
\bibitem [{\citenamefont {Ganzhorn}\ \emph {et~al.}(2013)\citenamefont {Ganzhorn}, \citenamefont {Klyatskaya}, \citenamefont {Ruben},\ and\ \citenamefont {Wernsdorfer}}]{Ganzhorn2013}%
  \BibitemOpen
  \bibfield  {author} {\bibinfo {author} {\bibfnamefont {M.}~\bibnamefont {Ganzhorn}}, \bibinfo {author} {\bibfnamefont {S.}~\bibnamefont {Klyatskaya}}, \bibinfo {author} {\bibfnamefont {M.}~\bibnamefont {Ruben}},\ and\ \bibinfo {author} {\bibfnamefont {W.}~\bibnamefont {Wernsdorfer}},\ }\bibfield  {title} {\bibinfo {title} {Strong spin–phonon coupling between a single-molecule magnet and a carbon nanotube nanoelectromechanical system},\ }\href {https://doi.org/10.1038/nnano.2012.258} {\bibfield  {journal} {\bibinfo  {journal} {Nature Nanotechnology}\ }\textbf {\bibinfo {volume} {8}},\ \bibinfo {pages} {165} (\bibinfo {year} {2013})}\BibitemShut {NoStop}%
\bibitem [{\citenamefont {Vincent}\ \emph {et~al.}(2012)\citenamefont {Vincent}, \citenamefont {Klyatskaya}, \citenamefont {Ruben}, \citenamefont {Wernsdorfer},\ and\ \citenamefont {Balestro}}]{Vincent2012}%
  \BibitemOpen
  \bibfield  {author} {\bibinfo {author} {\bibfnamefont {R.}~\bibnamefont {Vincent}}, \bibinfo {author} {\bibfnamefont {S.}~\bibnamefont {Klyatskaya}}, \bibinfo {author} {\bibfnamefont {M.}~\bibnamefont {Ruben}}, \bibinfo {author} {\bibfnamefont {W.}~\bibnamefont {Wernsdorfer}},\ and\ \bibinfo {author} {\bibfnamefont {F.}~\bibnamefont {Balestro}},\ }\bibfield  {title} {\bibinfo {title} {Electronic read-out of a single nuclear spin using a molecular spin transistor},\ }\href {https://doi.org/10.1038/nature11341} {\bibfield  {journal} {\bibinfo  {journal} {Nature}\ }\textbf {\bibinfo {volume} {488}},\ \bibinfo {pages} {357} (\bibinfo {year} {2012})}\BibitemShut {NoStop}%
\bibitem [{\citenamefont {Lumetti}\ \emph {et~al.}(2016)\citenamefont {Lumetti}, \citenamefont {Candini}, \citenamefont {Godfrin}, \citenamefont {Balestro}, \citenamefont {Wernsdorfer}, \citenamefont {Klyatskaya}, \citenamefont {Ruben},\ and\ \citenamefont {Affronte}}]{Lumetti2016}%
  \BibitemOpen
  \bibfield  {author} {\bibinfo {author} {\bibfnamefont {S.}~\bibnamefont {Lumetti}}, \bibinfo {author} {\bibfnamefont {A.}~\bibnamefont {Candini}}, \bibinfo {author} {\bibfnamefont {C.}~\bibnamefont {Godfrin}}, \bibinfo {author} {\bibfnamefont {F.}~\bibnamefont {Balestro}}, \bibinfo {author} {\bibfnamefont {W.}~\bibnamefont {Wernsdorfer}}, \bibinfo {author} {\bibfnamefont {S.}~\bibnamefont {Klyatskaya}}, \bibinfo {author} {\bibfnamefont {M.}~\bibnamefont {Ruben}},\ and\ \bibinfo {author} {\bibfnamefont {M.}~\bibnamefont {Affronte}},\ }\bibfield  {title} {\bibinfo {title} {Single-molecule devices with graphene electrodes},\ }\href {https://doi.org/10.1039/C6DT02445A} {\bibfield  {journal} {\bibinfo  {journal} {Dalton Transactions}\ }\textbf {\bibinfo {volume} {45}},\ \bibinfo {pages} {16570} (\bibinfo {year} {2016})}\BibitemShut {NoStop}%
\bibitem [{\citenamefont {Jock}\ \emph {et~al.}(2018)\citenamefont {Jock}, \citenamefont {Jacobson}, \citenamefont {Harvey-Collard}, \citenamefont {Mounce}, \citenamefont {Srinivasa}, \citenamefont {Ward}, \citenamefont {Anderson}, \citenamefont {Manginell}, \citenamefont {Wendt}, \citenamefont {Rudolph}, \citenamefont {Pluym}, \citenamefont {Gamble}, \citenamefont {Baczewski}, \citenamefont {Witzel},\ and\ \citenamefont {Carroll}}]{Jock2018}%
  \BibitemOpen
  \bibfield  {author} {\bibinfo {author} {\bibfnamefont {R.~M.}\ \bibnamefont {Jock}}, \bibinfo {author} {\bibfnamefont {N.~T.}\ \bibnamefont {Jacobson}}, \bibinfo {author} {\bibfnamefont {P.}~\bibnamefont {Harvey-Collard}}, \bibinfo {author} {\bibfnamefont {A.~M.}\ \bibnamefont {Mounce}}, \bibinfo {author} {\bibfnamefont {V.}~\bibnamefont {Srinivasa}}, \bibinfo {author} {\bibfnamefont {D.~R.}\ \bibnamefont {Ward}}, \bibinfo {author} {\bibfnamefont {J.}~\bibnamefont {Anderson}}, \bibinfo {author} {\bibfnamefont {R.}~\bibnamefont {Manginell}}, \bibinfo {author} {\bibfnamefont {J.~R.}\ \bibnamefont {Wendt}}, \bibinfo {author} {\bibfnamefont {M.}~\bibnamefont {Rudolph}}, \bibinfo {author} {\bibfnamefont {T.}~\bibnamefont {Pluym}}, \bibinfo {author} {\bibfnamefont {J.~K.}\ \bibnamefont {Gamble}}, \bibinfo {author} {\bibfnamefont {A.~D.}\ \bibnamefont {Baczewski}}, \bibinfo {author} {\bibfnamefont {W.~M.}\ \bibnamefont {Witzel}},\ and\ \bibinfo {author} {\bibfnamefont {M.~S.}\ \bibnamefont {Carroll}},\ }\bibfield
  {title} {\bibinfo {title} {A silicon metal-oxide-semiconductor electron spin-orbit qubit},\ }\href {https://doi.org/10.1038/s41467-018-04200-0} {\bibfield  {journal} {\bibinfo  {journal} {Nature Communications}\ }\textbf {\bibinfo {volume} {9}},\ \bibinfo {pages} {1768} (\bibinfo {year} {2018})}\BibitemShut {NoStop}%
\bibitem [{\citenamefont {Steinacker}\ \emph {et~al.}(2024)\citenamefont {Steinacker}, \citenamefont {Stuyck}, \citenamefont {Lim}, \citenamefont {Tanttu}, \citenamefont {Feng}, \citenamefont {Nickl}, \citenamefont {Serrano}, \citenamefont {Candido}, \citenamefont {Cifuentes}, \citenamefont {Hudson}, \citenamefont {Chan}, \citenamefont {Kubicek}, \citenamefont {Jussot}, \citenamefont {Canvel}, \citenamefont {Beyne}, \citenamefont {Shimura}, \citenamefont {Loo}, \citenamefont {Godfrin}, \citenamefont {Raes}, \citenamefont {Baudot}, \citenamefont {Wan}, \citenamefont {Laucht}, \citenamefont {Yang}, \citenamefont {Saraiva}, \citenamefont {Escott}, \citenamefont {Greve},\ and\ \citenamefont {Dzurak}}]{Steinacker2024}%
  \BibitemOpen
  \bibfield  {author} {\bibinfo {author} {\bibfnamefont {P.}~\bibnamefont {Steinacker}}, \bibinfo {author} {\bibfnamefont {N.~D.}\ \bibnamefont {Stuyck}}, \bibinfo {author} {\bibfnamefont {W.~H.}\ \bibnamefont {Lim}}, \bibinfo {author} {\bibfnamefont {T.}~\bibnamefont {Tanttu}}, \bibinfo {author} {\bibfnamefont {M.}~\bibnamefont {Feng}}, \bibinfo {author} {\bibfnamefont {A.}~\bibnamefont {Nickl}}, \bibinfo {author} {\bibfnamefont {S.}~\bibnamefont {Serrano}}, \bibinfo {author} {\bibfnamefont {M.}~\bibnamefont {Candido}}, \bibinfo {author} {\bibfnamefont {J.~D.}\ \bibnamefont {Cifuentes}}, \bibinfo {author} {\bibfnamefont {F.~E.}\ \bibnamefont {Hudson}}, \bibinfo {author} {\bibfnamefont {K.~W.}\ \bibnamefont {Chan}}, \bibinfo {author} {\bibfnamefont {S.}~\bibnamefont {Kubicek}}, \bibinfo {author} {\bibfnamefont {J.}~\bibnamefont {Jussot}}, \bibinfo {author} {\bibfnamefont {Y.}~\bibnamefont {Canvel}}, \bibinfo {author} {\bibfnamefont {S.}~\bibnamefont {Beyne}}, \bibinfo {author} {\bibfnamefont {Y.}~\bibnamefont
  {Shimura}}, \bibinfo {author} {\bibfnamefont {R.}~\bibnamefont {Loo}}, \bibinfo {author} {\bibfnamefont {C.}~\bibnamefont {Godfrin}}, \bibinfo {author} {\bibfnamefont {B.}~\bibnamefont {Raes}}, \bibinfo {author} {\bibfnamefont {S.}~\bibnamefont {Baudot}}, \bibinfo {author} {\bibfnamefont {D.}~\bibnamefont {Wan}}, \bibinfo {author} {\bibfnamefont {A.}~\bibnamefont {Laucht}}, \bibinfo {author} {\bibfnamefont {C.~H.}\ \bibnamefont {Yang}}, \bibinfo {author} {\bibfnamefont {A.}~\bibnamefont {Saraiva}}, \bibinfo {author} {\bibfnamefont {C.~C.}\ \bibnamefont {Escott}}, \bibinfo {author} {\bibfnamefont {K.~D.}\ \bibnamefont {Greve}},\ and\ \bibinfo {author} {\bibfnamefont {A.~S.}\ \bibnamefont {Dzurak}},\ }\bibfield  {title} {\bibinfo {title} {A 300 mm foundry silicon spin qubit unit cell exceeding 99\% fidelity in all operations},\ }\href@noop {} {\bibfield  {journal} {\bibinfo  {journal} {arXiv preprint}\ } (\bibinfo {year} {2024})},\ \bibinfo {note} {submitted 21 October 2024},\ \Eprint
  {https://arxiv.org/abs/2410.15590} {arXiv:2410.15590 [cond-mat.mes-hall]} \BibitemShut {NoStop}%
\bibitem [{\citenamefont {Elzerman}\ \emph {et~al.}(2004)\citenamefont {Elzerman}, \citenamefont {Hanson}, \citenamefont {van Beveren}, \citenamefont {Witkamp}, \citenamefont {Vandersypen},\ and\ \citenamefont {Kouwenhoven}}]{Elzerman2004}%
  \BibitemOpen
  \bibfield  {author} {\bibinfo {author} {\bibfnamefont {J.~M.}\ \bibnamefont {Elzerman}}, \bibinfo {author} {\bibfnamefont {R.}~\bibnamefont {Hanson}}, \bibinfo {author} {\bibfnamefont {L.~H.~W.}\ \bibnamefont {van Beveren}}, \bibinfo {author} {\bibfnamefont {B.}~\bibnamefont {Witkamp}}, \bibinfo {author} {\bibfnamefont {L.~M.~K.}\ \bibnamefont {Vandersypen}},\ and\ \bibinfo {author} {\bibfnamefont {L.~P.}\ \bibnamefont {Kouwenhoven}},\ }\bibfield  {title} {\bibinfo {title} {Single-shot read-out of an individual electron spin in a quantum dot},\ }\href {https://doi.org/10.1038/nature02693} {\bibfield  {journal} {\bibinfo  {journal} {Nature}\ }\textbf {\bibinfo {volume} {430}},\ \bibinfo {pages} {431} (\bibinfo {year} {2004})}\BibitemShut {NoStop}%
\bibitem [{\citenamefont {Hu}\ \emph {et~al.}(2021)\citenamefont {Hu}, \citenamefont {Ma}, \citenamefont {Ni}, \citenamefont {Zhang}, \citenamefont {Zhou}, \citenamefont {Wang}, \citenamefont {Luo}, \citenamefont {Cao}, \citenamefont {Kong}, \citenamefont {Wang}, \citenamefont {Li},\ and\ \citenamefont {Guo}}]{Hu2021}%
  \BibitemOpen
  \bibfield  {author} {\bibinfo {author} {\bibfnamefont {R.}~\bibnamefont {Hu}}, \bibinfo {author} {\bibfnamefont {R.}~\bibnamefont {Ma}}, \bibinfo {author} {\bibfnamefont {M.}~\bibnamefont {Ni}}, \bibinfo {author} {\bibfnamefont {X.}~\bibnamefont {Zhang}}, \bibinfo {author} {\bibfnamefont {Y.}~\bibnamefont {Zhou}}, \bibinfo {author} {\bibfnamefont {K.}~\bibnamefont {Wang}}, \bibinfo {author} {\bibfnamefont {G.}~\bibnamefont {Luo}}, \bibinfo {author} {\bibfnamefont {G.}~\bibnamefont {Cao}}, \bibinfo {author} {\bibfnamefont {Z.}~\bibnamefont {Kong}}, \bibinfo {author} {\bibfnamefont {G.}~\bibnamefont {Wang}}, \bibinfo {author} {\bibfnamefont {H.}~\bibnamefont {Li}},\ and\ \bibinfo {author} {\bibfnamefont {G.}~\bibnamefont {Guo}},\ }\bibfield  {title} {\bibinfo {title} {An operation guide of si‑mos quantum dots for spin qubits},\ }\href {https://doi.org/10.3390/nano11102486} {\bibfield  {journal} {\bibinfo  {journal} {Nanomaterials}\ }\textbf {\bibinfo {volume} {11}},\ \bibinfo {pages} {2486} (\bibinfo {year}
  {2021})}\BibitemShut {NoStop}%
\bibitem [{\citenamefont {Veldhorst}\ \emph {et~al.}(2015)\citenamefont {Veldhorst}, \citenamefont {Yang}, \citenamefont {Hwang}, \citenamefont {Huang}, \citenamefont {Dehollain}, \citenamefont {Muhonen}, \citenamefont {Simmons}, \citenamefont {Laucht}, \citenamefont {Hudson}, \citenamefont {Itoh}, \citenamefont {Morello},\ and\ \citenamefont {Dzurak}}]{Veldhorst2015}%
  \BibitemOpen
  \bibfield  {author} {\bibinfo {author} {\bibfnamefont {M.}~\bibnamefont {Veldhorst}}, \bibinfo {author} {\bibfnamefont {C.~H.}\ \bibnamefont {Yang}}, \bibinfo {author} {\bibfnamefont {J.~C.~C.}\ \bibnamefont {Hwang}}, \bibinfo {author} {\bibfnamefont {W.}~\bibnamefont {Huang}}, \bibinfo {author} {\bibfnamefont {J.~P.}\ \bibnamefont {Dehollain}}, \bibinfo {author} {\bibfnamefont {J.~T.}\ \bibnamefont {Muhonen}}, \bibinfo {author} {\bibfnamefont {S.}~\bibnamefont {Simmons}}, \bibinfo {author} {\bibfnamefont {A.}~\bibnamefont {Laucht}}, \bibinfo {author} {\bibfnamefont {F.~E.}\ \bibnamefont {Hudson}}, \bibinfo {author} {\bibfnamefont {K.~M.}\ \bibnamefont {Itoh}}, \bibinfo {author} {\bibfnamefont {A.}~\bibnamefont {Morello}},\ and\ \bibinfo {author} {\bibfnamefont {A.~S.}\ \bibnamefont {Dzurak}},\ }\bibfield  {title} {\bibinfo {title} {A two-qubit logic gate in silicon},\ }\href {https://doi.org/10.1038/nature15263} {\bibfield  {journal} {\bibinfo  {journal} {Nature}\ }\textbf {\bibinfo {volume} {526}},\
  \bibinfo {pages} {410} (\bibinfo {year} {2015})}\BibitemShut {NoStop}%
\bibitem [{\citenamefont {Zhao}\ \emph {et~al.}(2019)\citenamefont {Zhao}, \citenamefont {Tanttu}, \citenamefont {Tan}, \citenamefont {Hensen}, \citenamefont {Chan}, \citenamefont {Hwang}, \citenamefont {Leon}, \citenamefont {Yang}, \citenamefont {Gilbert}, \citenamefont {Hudson}, \citenamefont {Itoh}, \citenamefont {Kiselev}, \citenamefont {Ladd}, \citenamefont {Morello}, \citenamefont {Laucht},\ and\ \citenamefont {Dzurak}}]{Zhao2019}%
  \BibitemOpen
  \bibfield  {author} {\bibinfo {author} {\bibfnamefont {R.}~\bibnamefont {Zhao}}, \bibinfo {author} {\bibfnamefont {T.}~\bibnamefont {Tanttu}}, \bibinfo {author} {\bibfnamefont {K.~Y.}\ \bibnamefont {Tan}}, \bibinfo {author} {\bibfnamefont {B.}~\bibnamefont {Hensen}}, \bibinfo {author} {\bibfnamefont {K.~W.}\ \bibnamefont {Chan}}, \bibinfo {author} {\bibfnamefont {J.~C.}\ \bibnamefont {Hwang}}, \bibinfo {author} {\bibfnamefont {R.~C.~C.}\ \bibnamefont {Leon}}, \bibinfo {author} {\bibfnamefont {C.~H.}\ \bibnamefont {Yang}}, \bibinfo {author} {\bibfnamefont {W.}~\bibnamefont {Gilbert}}, \bibinfo {author} {\bibfnamefont {F.~E.}\ \bibnamefont {Hudson}}, \bibinfo {author} {\bibfnamefont {K.~M.}\ \bibnamefont {Itoh}}, \bibinfo {author} {\bibfnamefont {A.~A.}\ \bibnamefont {Kiselev}}, \bibinfo {author} {\bibfnamefont {T.~D.}\ \bibnamefont {Ladd}}, \bibinfo {author} {\bibfnamefont {A.}~\bibnamefont {Morello}}, \bibinfo {author} {\bibfnamefont {A.}~\bibnamefont {Laucht}},\ and\ \bibinfo {author} {\bibfnamefont
  {A.~S.}\ \bibnamefont {Dzurak}},\ }\bibfield  {title} {\bibinfo {title} {Single-spin qubits in isotopically enriched silicon at low magnetic field},\ }\href {https://doi.org/10.1038/s41467-019-13416-7} {\bibfield  {journal} {\bibinfo  {journal} {Nature Communications}\ }\textbf {\bibinfo {volume} {10}},\ \bibinfo {pages} {5500} (\bibinfo {year} {2019})}\BibitemShut {NoStop}%
\bibitem [{\citenamefont {Vahapoglu}\ \emph {et~al.}(2022)\citenamefont {Vahapoglu}, \citenamefont {Slack-Smith}, \citenamefont {Leon}, \citenamefont {Lim}, \citenamefont {Hudson}, \citenamefont {Day}, \citenamefont {Cifuentes}, \citenamefont {Tanttu}, \citenamefont {Yang}, \citenamefont {Saraiva}, \citenamefont {Abrosimov}, \citenamefont {Pohl}, \citenamefont {Thewalt}, \citenamefont {Laucht}, \citenamefont {Dzurak},\ and\ \citenamefont {Pla}}]{Vahapoglu2022}%
  \BibitemOpen
  \bibfield  {author} {\bibinfo {author} {\bibfnamefont {E.}~\bibnamefont {Vahapoglu}}, \bibinfo {author} {\bibfnamefont {J.~P.}\ \bibnamefont {Slack-Smith}}, \bibinfo {author} {\bibfnamefont {R.~C.~C.}\ \bibnamefont {Leon}}, \bibinfo {author} {\bibfnamefont {W.~H.}\ \bibnamefont {Lim}}, \bibinfo {author} {\bibfnamefont {F.~E.}\ \bibnamefont {Hudson}}, \bibinfo {author} {\bibfnamefont {T.}~\bibnamefont {Day}}, \bibinfo {author} {\bibfnamefont {J.~D.}\ \bibnamefont {Cifuentes}}, \bibinfo {author} {\bibfnamefont {T.}~\bibnamefont {Tanttu}}, \bibinfo {author} {\bibfnamefont {C.~H.}\ \bibnamefont {Yang}}, \bibinfo {author} {\bibfnamefont {A.}~\bibnamefont {Saraiva}}, \bibinfo {author} {\bibfnamefont {N.~V.}\ \bibnamefont {Abrosimov}}, \bibinfo {author} {\bibfnamefont {H.-J.}\ \bibnamefont {Pohl}}, \bibinfo {author} {\bibfnamefont {M.~L.~W.}\ \bibnamefont {Thewalt}}, \bibinfo {author} {\bibfnamefont {A.}~\bibnamefont {Laucht}}, \bibinfo {author} {\bibfnamefont {A.~S.}\ \bibnamefont {Dzurak}},\ and\ \bibinfo
  {author} {\bibfnamefont {J.~J.}\ \bibnamefont {Pla}},\ }\bibfield  {title} {\bibinfo {title} {Coherent control of electron spin qubits in silicon using a global field},\ }\href {https://doi.org/10.1038/s41534-022-00645-w} {\bibfield  {journal} {\bibinfo  {journal} {npj Quantum Information}\ }\textbf {\bibinfo {volume} {8}},\ \bibinfo {pages} {126} (\bibinfo {year} {2022})}\BibitemShut {NoStop}%
\bibitem [{\citenamefont {Stano}\ and\ \citenamefont {Loss}(2022)}]{Stano2022}%
  \BibitemOpen
  \bibfield  {author} {\bibinfo {author} {\bibfnamefont {P.}~\bibnamefont {Stano}}\ and\ \bibinfo {author} {\bibfnamefont {D.}~\bibnamefont {Loss}},\ }\bibfield  {title} {\bibinfo {title} {Review of performance metrics of spin qubits in gated semiconducting nanostructures},\ }\href {https://doi.org/10.1038/s42254-022-00484-w} {\bibfield  {journal} {\bibinfo  {journal} {Nature Reviews Physics}\ }\textbf {\bibinfo {volume} {4}},\ \bibinfo {pages} {672} (\bibinfo {year} {2022})}\BibitemShut {NoStop}%
\bibitem [{\citenamefont {Peña}\ \emph {et~al.}(2024)\citenamefont {Peña}, \citenamefont {Koepke}, \citenamefont {Dycus}, \citenamefont {Mounce}, \citenamefont {Baczewski}, \citenamefont {Jacobson},\ and\ \citenamefont {Bussmann}}]{Pena2024}%
  \BibitemOpen
  \bibfield  {author} {\bibinfo {author} {\bibfnamefont {L.~F.}\ \bibnamefont {Peña}}, \bibinfo {author} {\bibfnamefont {J.~C.}\ \bibnamefont {Koepke}}, \bibinfo {author} {\bibfnamefont {J.~H.}\ \bibnamefont {Dycus}}, \bibinfo {author} {\bibfnamefont {A.}~\bibnamefont {Mounce}}, \bibinfo {author} {\bibfnamefont {A.~D.}\ \bibnamefont {Baczewski}}, \bibinfo {author} {\bibfnamefont {N.~T.}\ \bibnamefont {Jacobson}},\ and\ \bibinfo {author} {\bibfnamefont {E.}~\bibnamefont {Bussmann}},\ }\bibfield  {title} {\bibinfo {title} {Modeling si/sige quantum dot variability induced by interface disorder reconstructed from multiperspective microscopy},\ }\href {https://doi.org/10.1038/s41534-024-00827-8} {\bibfield  {journal} {\bibinfo  {journal} {npj Quantum Information}\ }\textbf {\bibinfo {volume} {10}},\ \bibinfo {pages} {33} (\bibinfo {year} {2024})}\BibitemShut {NoStop}%
\bibitem [{\citenamefont {Koch}\ \emph {et~al.}(2025)\citenamefont {Koch}, \citenamefont {Godfrin}, \citenamefont {Adam}, \citenamefont {Ferrero}, \citenamefont {Schroller}, \citenamefont {Glaeser}, \citenamefont {Kubicek}, \citenamefont {Li}, \citenamefont {Loo}, \citenamefont {Massar}, \citenamefont {Simion}, \citenamefont {Wan}, \citenamefont {Greve},\ and\ \citenamefont {Wernsdorfer}}]{Koch2025}%
  \BibitemOpen
  \bibfield  {author} {\bibinfo {author} {\bibfnamefont {T.}~\bibnamefont {Koch}}, \bibinfo {author} {\bibfnamefont {C.}~\bibnamefont {Godfrin}}, \bibinfo {author} {\bibfnamefont {V.}~\bibnamefont {Adam}}, \bibinfo {author} {\bibfnamefont {J.}~\bibnamefont {Ferrero}}, \bibinfo {author} {\bibfnamefont {D.}~\bibnamefont {Schroller}}, \bibinfo {author} {\bibfnamefont {N.}~\bibnamefont {Glaeser}}, \bibinfo {author} {\bibfnamefont {S.}~\bibnamefont {Kubicek}}, \bibinfo {author} {\bibfnamefont {R.}~\bibnamefont {Li}}, \bibinfo {author} {\bibfnamefont {R.}~\bibnamefont {Loo}}, \bibinfo {author} {\bibfnamefont {S.}~\bibnamefont {Massar}}, \bibinfo {author} {\bibfnamefont {G.}~\bibnamefont {Simion}}, \bibinfo {author} {\bibfnamefont {D.}~\bibnamefont {Wan}}, \bibinfo {author} {\bibfnamefont {K.~D.}\ \bibnamefont {Greve}},\ and\ \bibinfo {author} {\bibfnamefont {W.}~\bibnamefont {Wernsdorfer}},\ }\bibfield  {title} {\bibinfo {title} {Industrial 300\,mm wafer processed spin qubits in natural
  silicon/silicon–germanium},\ }\bibfield  {journal} {\bibinfo  {journal} {npj Quantum Information}\ }\textbf {\bibinfo {volume} {11}},\ \href {https://doi.org/10.1038/s41534-025-01016-x} {10.1038/s41534-025-01016-x} (\bibinfo {year} {2025})\BibitemShut {NoStop}%
\bibitem [{\citenamefont {van Riggelen‑Doelman}\ \emph {et~al.}(2024)\citenamefont {van Riggelen‑Doelman}, \citenamefont {Wang}, \citenamefont {de~Snoo}, \citenamefont {Lawrie}, \citenamefont {Hendrickx}, \citenamefont {Rimbach‑Russ}, \citenamefont {Sammak}, \citenamefont {Scappucci}, \citenamefont {Déprez},\ and\ \citenamefont {Veldhorst}}]{vanRiggelen-Doelman2024}%
  \BibitemOpen
  \bibfield  {author} {\bibinfo {author} {\bibfnamefont {F.}~\bibnamefont {van Riggelen‑Doelman}}, \bibinfo {author} {\bibfnamefont {C.}~\bibnamefont {Wang}}, \bibinfo {author} {\bibfnamefont {S.~L.}\ \bibnamefont {de~Snoo}}, \bibinfo {author} {\bibfnamefont {W.~I.~L.}\ \bibnamefont {Lawrie}}, \bibinfo {author} {\bibfnamefont {N.~W.}\ \bibnamefont {Hendrickx}}, \bibinfo {author} {\bibfnamefont {M.}~\bibnamefont {Rimbach‑Russ}}, \bibinfo {author} {\bibfnamefont {A.}~\bibnamefont {Sammak}}, \bibinfo {author} {\bibfnamefont {G.}~\bibnamefont {Scappucci}}, \bibinfo {author} {\bibfnamefont {C.}~\bibnamefont {Déprez}},\ and\ \bibinfo {author} {\bibfnamefont {M.}~\bibnamefont {Veldhorst}},\ }\bibfield  {title} {\bibinfo {title} {Coherent spin qubit shuttling through germanium quantum dots},\ }\href {https://doi.org/10.1038/s41467-024-49358-y} {\bibfield  {journal} {\bibinfo  {journal} {Nature Communications}\ }\textbf {\bibinfo {volume} {15}},\ \bibinfo {pages} {5716} (\bibinfo {year} {2024})}\BibitemShut
  {NoStop}%
\bibitem [{\citenamefont {Koppens}\ \emph {et~al.}(2006)\citenamefont {Koppens}, \citenamefont {Buizert}, \citenamefont {Tielrooij}, \citenamefont {Vink}, \citenamefont {Nowack}, \citenamefont {Meunier}, \citenamefont {Kouwenhoven},\ and\ \citenamefont {Vandersypen}}]{Koppens2006}%
  \BibitemOpen
  \bibfield  {author} {\bibinfo {author} {\bibfnamefont {F.~H.~L.}\ \bibnamefont {Koppens}}, \bibinfo {author} {\bibfnamefont {C.}~\bibnamefont {Buizert}}, \bibinfo {author} {\bibfnamefont {K.~J.}\ \bibnamefont {Tielrooij}}, \bibinfo {author} {\bibfnamefont {I.~T.}\ \bibnamefont {Vink}}, \bibinfo {author} {\bibfnamefont {K.~C.}\ \bibnamefont {Nowack}}, \bibinfo {author} {\bibfnamefont {T.}~\bibnamefont {Meunier}}, \bibinfo {author} {\bibfnamefont {L.~P.}\ \bibnamefont {Kouwenhoven}},\ and\ \bibinfo {author} {\bibfnamefont {L.~M.~K.}\ \bibnamefont {Vandersypen}},\ }\bibfield  {title} {\bibinfo {title} {Driven coherent oscillations of a single electron spin in a quantum dot},\ }\href {https://doi.org/10.1038/nature05065} {\bibfield  {journal} {\bibinfo  {journal} {Nature}\ }\textbf {\bibinfo {volume} {442}},\ \bibinfo {pages} {766} (\bibinfo {year} {2006})}\BibitemShut {NoStop}%
\bibitem [{\citenamefont {Kittel}(2005)}]{Kittel2005}%
  \BibitemOpen
  \bibfield  {author} {\bibinfo {author} {\bibfnamefont {C.}~\bibnamefont {Kittel}},\ }\href@noop {} {\emph {\bibinfo {title} {Introduction to Solid State Physics}}},\ \bibinfo {edition} {8th}\ ed.\ (\bibinfo  {publisher} {Wiley},\ \bibinfo {year} {2005})\BibitemShut {NoStop}%
\bibitem [{\citenamefont {Jackson}(1999)}]{Jackson1999}%
  \BibitemOpen
  \bibfield  {author} {\bibinfo {author} {\bibfnamefont {J.~D.}\ \bibnamefont {Jackson}},\ }\href@noop {} {\emph {\bibinfo {title} {Classical Electrodynamics}}},\ \bibinfo {edition} {3rd}\ ed.\ (\bibinfo  {publisher} {Wiley},\ \bibinfo {year} {1999})\BibitemShut {NoStop}%
\bibitem [{\citenamefont {Hwang}\ \emph {et~al.}(2017)\citenamefont {Hwang}, \citenamefont {Yang}, \citenamefont {Hwang} \emph {et~al.}}]{Hwang2017}%
  \BibitemOpen
  \bibfield  {author} {\bibinfo {author} {\bibfnamefont {J.~C.~C.}\ \bibnamefont {Hwang}}, \bibinfo {author} {\bibfnamefont {C.~H.}\ \bibnamefont {Yang}}, \bibinfo {author} {\bibfnamefont {J.~C.~C.}\ \bibnamefont {Hwang}}, \emph {et~al.},\ }\bibfield  {title} {\bibinfo {title} {Impact of g-factors and valleys on spin qubits in a silicon metal-oxide-semiconductor double quantum dot},\ }\href {https://doi.org/10.1103/PhysRevB.96.045302} {\bibfield  {journal} {\bibinfo  {journal} {Physical Review B}\ }\textbf {\bibinfo {volume} {96}},\ \bibinfo {pages} {045302} (\bibinfo {year} {2017})}\BibitemShut {NoStop}%
\bibitem [{\citenamefont {Abragam}(1961)}]{abragam1961}%
  \BibitemOpen
  \bibfield  {author} {\bibinfo {author} {\bibfnamefont {A.}~\bibnamefont {Abragam}},\ }\href@noop {} {\emph {\bibinfo {title} {The Principles of Nuclear Magnetism}}},\ International Series of Monographs on Physics\ (\bibinfo  {publisher} {Clarendon Press},\ \bibinfo {address} {Oxford},\ \bibinfo {year} {1961})\BibitemShut {NoStop}%
\bibitem [{\citenamefont {Harris}\ and\ \citenamefont {Hoover}(1972)}]{Harris1972}%
  \BibitemOpen
  \bibfield  {author} {\bibinfo {author} {\bibfnamefont {C.~B.}\ \bibnamefont {Harris}}\ and\ \bibinfo {author} {\bibfnamefont {R.~J.}\ \bibnamefont {Hoover}},\ }\bibfield  {title} {\bibinfo {title} {Optically detected adiabatic inversion in phosphorescent triplet states and the measurement of intramolecular energy transfer processes},\ }\href {https://doi.org/10.1063/1.1677520} {\bibfield  {journal} {\bibinfo  {journal} {The Journal of Chemical Physics}\ }\textbf {\bibinfo {volume} {56}},\ \bibinfo {pages} {2199} (\bibinfo {year} {1972})}\BibitemShut {NoStop}%
\bibitem [{\citenamefont {Li}\ \emph {et~al.}(2023)\citenamefont {Li}, \citenamefont {Spierings},\ and\ \citenamefont {Steinberg}}]{Li2023}%
  \BibitemOpen
  \bibfield  {author} {\bibinfo {author} {\bibfnamefont {K.}~\bibnamefont {Li}}, \bibinfo {author} {\bibfnamefont {D.~C.}\ \bibnamefont {Spierings}},\ and\ \bibinfo {author} {\bibfnamefont {A.~M.}\ \bibnamefont {Steinberg}},\ }\bibfield  {title} {\bibinfo {title} {Efficient adiabatic rapid passage in the presence of noise},\ }\href {https://doi.org/10.1103/PhysRevA.108.012615} {\bibfield  {journal} {\bibinfo  {journal} {Physical Review A}\ }\textbf {\bibinfo {volume} {108}},\ \bibinfo {pages} {012615} (\bibinfo {year} {2023})}\BibitemShut {NoStop}%
\bibitem [{\citenamefont {Fogarty}\ \emph {et~al.}(2018)\citenamefont {Fogarty}, \citenamefont {Chan}, \citenamefont {Hensen}, \citenamefont {Huang}, \citenamefont {Tanttu}, \citenamefont {Yang}, \citenamefont {Laucht}, \citenamefont {Veldhorst}, \citenamefont {Hudson}, \citenamefont {Itoh}, \citenamefont {Culcer}, \citenamefont {Ladd}, \citenamefont {Morello},\ and\ \citenamefont {Dzurak}}]{Fogarty2018}%
  \BibitemOpen
  \bibfield  {author} {\bibinfo {author} {\bibfnamefont {M.~A.}\ \bibnamefont {Fogarty}}, \bibinfo {author} {\bibfnamefont {K.~W.}\ \bibnamefont {Chan}}, \bibinfo {author} {\bibfnamefont {B.}~\bibnamefont {Hensen}}, \bibinfo {author} {\bibfnamefont {W.}~\bibnamefont {Huang}}, \bibinfo {author} {\bibfnamefont {T.}~\bibnamefont {Tanttu}}, \bibinfo {author} {\bibfnamefont {C.~H.}\ \bibnamefont {Yang}}, \bibinfo {author} {\bibfnamefont {A.}~\bibnamefont {Laucht}}, \bibinfo {author} {\bibfnamefont {M.}~\bibnamefont {Veldhorst}}, \bibinfo {author} {\bibfnamefont {F.~E.}\ \bibnamefont {Hudson}}, \bibinfo {author} {\bibfnamefont {K.~M.}\ \bibnamefont {Itoh}}, \bibinfo {author} {\bibfnamefont {D.}~\bibnamefont {Culcer}}, \bibinfo {author} {\bibfnamefont {T.~D.}\ \bibnamefont {Ladd}}, \bibinfo {author} {\bibfnamefont {A.}~\bibnamefont {Morello}},\ and\ \bibinfo {author} {\bibfnamefont {A.~S.}\ \bibnamefont {Dzurak}},\ }\bibfield  {title} {\bibinfo {title} {Integrated silicon qubit platform with single-spin
  addressability, exchange control and single-shot singlet-triplet readout},\ }\href {https://doi.org/10.1038/s41467-018-06039-x} {\bibfield  {journal} {\bibinfo  {journal} {Nature Communications}\ }\textbf {\bibinfo {volume} {9}},\ \bibinfo {pages} {4370} (\bibinfo {year} {2018})}\BibitemShut {NoStop}%
\bibitem [{\citenamefont {Perfetti}\ \emph {et~al.}(2016)\citenamefont {Perfetti}, \citenamefont {Serri}, \citenamefont {Poggini}, \citenamefont {Mannini}, \citenamefont {Rovai}, \citenamefont {Sainctavit}, \citenamefont {Heutz},\ and\ \citenamefont {Sessoli}}]{Perfetti2016}%
  \BibitemOpen
  \bibfield  {author} {\bibinfo {author} {\bibfnamefont {M.}~\bibnamefont {Perfetti}}, \bibinfo {author} {\bibfnamefont {M.}~\bibnamefont {Serri}}, \bibinfo {author} {\bibfnamefont {L.}~\bibnamefont {Poggini}}, \bibinfo {author} {\bibfnamefont {M.}~\bibnamefont {Mannini}}, \bibinfo {author} {\bibfnamefont {D.}~\bibnamefont {Rovai}}, \bibinfo {author} {\bibfnamefont {P.}~\bibnamefont {Sainctavit}}, \bibinfo {author} {\bibfnamefont {S.}~\bibnamefont {Heutz}},\ and\ \bibinfo {author} {\bibfnamefont {R.}~\bibnamefont {Sessoli}},\ }\bibfield  {title} {\bibinfo {title} {Molecular order in buried layers of tbpc$_2$ single-molecule magnets detected by torque magnetometry},\ }\href {https://doi.org/10.1002/adma.201600791} {\bibfield  {journal} {\bibinfo  {journal} {Advanced Materials}\ }\textbf {\bibinfo {volume} {28}},\ \bibinfo {pages} {6946} (\bibinfo {year} {2016})}\BibitemShut {NoStop}%
\bibitem [{\citenamefont {Nabiałek}\ \emph {et~al.}(2010)\citenamefont {Nabiałek}, \citenamefont {Vasiliev}, \citenamefont {Chabanenko}, \citenamefont {Pérez‑Rodríguez}, \citenamefont {Piechota},\ and\ \citenamefont {Szymczak}}]{Nabialek2010}%
  \BibitemOpen
  \bibfield  {author} {\bibinfo {author} {\bibfnamefont {A.}~\bibnamefont {Nabiałek}}, \bibinfo {author} {\bibfnamefont {S.}~\bibnamefont {Vasiliev}}, \bibinfo {author} {\bibfnamefont {V.}~\bibnamefont {Chabanenko}}, \bibinfo {author} {\bibfnamefont {F.}~\bibnamefont {Pérez‑Rodríguez}}, \bibinfo {author} {\bibfnamefont {S.}~\bibnamefont {Piechota}},\ and\ \bibinfo {author} {\bibfnamefont {H.}~\bibnamefont {Szymczak}},\ }\bibfield  {title} {\bibinfo {title} {The influence of magnetic history on the stability of critical state and the dynamics of flux jumps in conventional nbti superconductor},\ }\href {https://doi.org/10.12693/APHYSPOLA.118.343} {\bibfield  {journal} {\bibinfo  {journal} {Acta Physica Polonica A}\ }\textbf {\bibinfo {volume} {118}},\ \bibinfo {pages} {343} (\bibinfo {year} {2010})}\BibitemShut {NoStop}%
\bibitem [{\citenamefont {Edholm}\ and\ \citenamefont {Blomberg}(2000)}]{Edholm2000}%
  \BibitemOpen
  \bibfield  {author} {\bibinfo {author} {\bibfnamefont {O.}~\bibnamefont {Edholm}}\ and\ \bibinfo {author} {\bibfnamefont {C.}~\bibnamefont {Blomberg}},\ }\bibfield  {title} {\bibinfo {title} {Stretched exponentials and barrier distributions},\ }\href {https://doi.org/10.1016/S0301-0104(99)00349-3} {\bibfield  {journal} {\bibinfo  {journal} {Chemical Physics}\ }\textbf {\bibinfo {volume} {252}},\ \bibinfo {pages} {221} (\bibinfo {year} {2000})}\BibitemShut {NoStop}%
\bibitem [{\citenamefont {Woodruff}\ \emph {et~al.}(2013)\citenamefont {Woodruff}, \citenamefont {Winpenny},\ and\ \citenamefont {Layfield}}]{Woodruff2013}%
  \BibitemOpen
  \bibfield  {author} {\bibinfo {author} {\bibfnamefont {D.~N.}\ \bibnamefont {Woodruff}}, \bibinfo {author} {\bibfnamefont {R.~E.~P.}\ \bibnamefont {Winpenny}},\ and\ \bibinfo {author} {\bibfnamefont {R.~A.}\ \bibnamefont {Layfield}},\ }\bibfield  {title} {\bibinfo {title} {Lanthanide single-molecule magnets},\ }\href {https://doi.org/10.1021/cr400018q} {\bibfield  {journal} {\bibinfo  {journal} {Chemical Reviews}\ }\textbf {\bibinfo {volume} {113}},\ \bibinfo {pages} {5110} (\bibinfo {year} {2013})}\BibitemShut {NoStop}%
\bibitem [{\citenamefont {Tosato}\ \emph {et~al.}(2025)\citenamefont {Tosato}, \citenamefont {Elsayed}, \citenamefont {Poggiali}, \citenamefont {Stehouwer}, \citenamefont {Costa}, \citenamefont {Hudson}, \citenamefont {Esposti},\ and\ \citenamefont {Scappucci}}]{Tosato2025}%
  \BibitemOpen
  \bibfield  {author} {\bibinfo {author} {\bibfnamefont {A.}~\bibnamefont {Tosato}}, \bibinfo {author} {\bibfnamefont {A.}~\bibnamefont {Elsayed}}, \bibinfo {author} {\bibfnamefont {F.}~\bibnamefont {Poggiali}}, \bibinfo {author} {\bibfnamefont {L.}~\bibnamefont {Stehouwer}}, \bibinfo {author} {\bibfnamefont {D.}~\bibnamefont {Costa}}, \bibinfo {author} {\bibfnamefont {K.}~\bibnamefont {Hudson}}, \bibinfo {author} {\bibfnamefont {D.~D.}\ \bibnamefont {Esposti}},\ and\ \bibinfo {author} {\bibfnamefont {G.}~\bibnamefont {Scappucci}},\ }\bibfield  {title} {\bibinfo {title} {Qarpet: A crossbar chip for benchmarking semiconductor spin qubits},\ }\href@noop {} {\bibfield  {journal} {\bibinfo  {journal} {arXiv preprint}\ } (\bibinfo {year} {2025})},\ \bibinfo {note} {published on arXiv 7 April 2025},\ \Eprint {https://arxiv.org/abs/2504.05460} {arXiv:2504.05460 [cond-mat.mes-hall]} \BibitemShut {NoStop}%
\bibitem [{\citenamefont {Smet}\ \emph {et~al.}(2025)\citenamefont {Smet}, \citenamefont {Matsumoto}, \citenamefont {Zwerver}, \citenamefont {Tryputen}, \citenamefont {de~Snoo}, \citenamefont {Amitonov}, \citenamefont {Katiraee-Far}, \citenamefont {Sammak}, \citenamefont {Samkharadze}, \citenamefont {Önder Gül}, \citenamefont {Wasserman}, \citenamefont {Greplová}, \citenamefont {Rimbach-Russ}, \citenamefont {Scappucci},\ and\ \citenamefont {Vandersypen}}]{DeSmet2025}%
  \BibitemOpen
  \bibfield  {author} {\bibinfo {author} {\bibfnamefont {M.~D.}\ \bibnamefont {Smet}}, \bibinfo {author} {\bibfnamefont {Y.}~\bibnamefont {Matsumoto}}, \bibinfo {author} {\bibfnamefont {A.-M.~J.}\ \bibnamefont {Zwerver}}, \bibinfo {author} {\bibfnamefont {L.}~\bibnamefont {Tryputen}}, \bibinfo {author} {\bibfnamefont {S.~L.}\ \bibnamefont {de~Snoo}}, \bibinfo {author} {\bibfnamefont {S.~V.}\ \bibnamefont {Amitonov}}, \bibinfo {author} {\bibfnamefont {S.~R.}\ \bibnamefont {Katiraee-Far}}, \bibinfo {author} {\bibfnamefont {A.}~\bibnamefont {Sammak}}, \bibinfo {author} {\bibfnamefont {N.}~\bibnamefont {Samkharadze}}, \bibinfo {author} {\bibnamefont {Önder Gül}}, \bibinfo {author} {\bibfnamefont {R.~N.~M.}\ \bibnamefont {Wasserman}}, \bibinfo {author} {\bibfnamefont {E.}~\bibnamefont {Greplová}}, \bibinfo {author} {\bibfnamefont {M.}~\bibnamefont {Rimbach-Russ}}, \bibinfo {author} {\bibfnamefont {G.}~\bibnamefont {Scappucci}},\ and\ \bibinfo {author} {\bibfnamefont {L.~M.~K.}\ \bibnamefont {Vandersypen}},\
  }\bibfield  {title} {\bibinfo {title} {High-fidelity single-spin shuttling in silicon},\ }\bibfield  {journal} {\bibinfo  {journal} {Nature Nanotechnology}\ }\href {https://doi.org/10.1038/s41565-025-01920-5} {10.1038/s41565-025-01920-5} (\bibinfo {year} {2025}),\ \bibinfo {note} {published online 9 June 2025}\BibitemShut {NoStop}%
\bibitem [{\citenamefont {Yamabayashi}\ \emph {et~al.}(2017)\citenamefont {Yamabayashi}, \citenamefont {Katoh}, \citenamefont {Breedlove},\ and\ \citenamefont {Yamashita}}]{Yamabayashi2017}%
  \BibitemOpen
  \bibfield  {author} {\bibinfo {author} {\bibfnamefont {T.}~\bibnamefont {Yamabayashi}}, \bibinfo {author} {\bibfnamefont {K.}~\bibnamefont {Katoh}}, \bibinfo {author} {\bibfnamefont {B.~K.}\ \bibnamefont {Breedlove}},\ and\ \bibinfo {author} {\bibfnamefont {M.}~\bibnamefont {Yamashita}},\ }\bibfield  {title} {\bibinfo {title} {Molecular orientation of a terbium(iii)-phthalocyaninato double-decker complex for effective suppression of quantum tunneling of the magnetization},\ }\href {https://doi.org/10.3390/molecules22060999} {\bibfield  {journal} {\bibinfo  {journal} {Molecules}\ }\textbf {\bibinfo {volume} {22}},\ \bibinfo {pages} {999} (\bibinfo {year} {2017})}\BibitemShut {NoStop}%
\bibitem [{\citenamefont {Malavolti}\ \emph {et~al.}(2013)\citenamefont {Malavolti}, \citenamefont {Mannini}, \citenamefont {Car}, \citenamefont {Campo}, \citenamefont {Pineider},\ and\ \citenamefont {Sessoli}}]{Malavolti2013}%
  \BibitemOpen
  \bibfield  {author} {\bibinfo {author} {\bibfnamefont {L.}~\bibnamefont {Malavolti}}, \bibinfo {author} {\bibfnamefont {M.}~\bibnamefont {Mannini}}, \bibinfo {author} {\bibfnamefont {P.-E.}\ \bibnamefont {Car}}, \bibinfo {author} {\bibfnamefont {G.}~\bibnamefont {Campo}}, \bibinfo {author} {\bibfnamefont {F.}~\bibnamefont {Pineider}},\ and\ \bibinfo {author} {\bibfnamefont {R.}~\bibnamefont {Sessoli}},\ }\bibfield  {title} {\bibinfo {title} {Erratic magnetic hysteresis of tbpc$_2$ molecular nanomagnets},\ }\href {https://doi.org/10.1039/C3TC00925D} {\bibfield  {journal} {\bibinfo  {journal} {Journal of Materials Chemistry C}\ }\textbf {\bibinfo {volume} {1}},\ \bibinfo {pages} {2935} (\bibinfo {year} {2013})}\BibitemShut {NoStop}%
\bibitem [{\citenamefont {Beenakker}(1991)}]{Beenakker1991}%
  \BibitemOpen
  \bibfield  {author} {\bibinfo {author} {\bibfnamefont {C.~W.~J.}\ \bibnamefont {Beenakker}},\ }\bibfield  {title} {\bibinfo {title} {Theory of coulomb-blockade oscillations in the conductance of a quantum dot},\ }\href {https://doi.org/10.1103/PhysRevB.44.1646} {\bibfield  {journal} {\bibinfo  {journal} {Physical Review B}\ }\textbf {\bibinfo {volume} {44}},\ \bibinfo {pages} {1646} (\bibinfo {year} {1991})}\BibitemShut {NoStop}%
\bibitem [{\citenamefont {Maradan}\ \emph {et~al.}(2014)\citenamefont {Maradan}, \citenamefont {Casparis}, \citenamefont {Liu}, \citenamefont {Biesinger}, \citenamefont {Scheller}, \citenamefont {Zumbühl}, \citenamefont {Zimmerman},\ and\ \citenamefont {Gossard}}]{Maradan2014}%
  \BibitemOpen
  \bibfield  {author} {\bibinfo {author} {\bibfnamefont {D.}~\bibnamefont {Maradan}}, \bibinfo {author} {\bibfnamefont {L.}~\bibnamefont {Casparis}}, \bibinfo {author} {\bibfnamefont {T.-M.}\ \bibnamefont {Liu}}, \bibinfo {author} {\bibfnamefont {D.~E.~F.}\ \bibnamefont {Biesinger}}, \bibinfo {author} {\bibfnamefont {C.~P.}\ \bibnamefont {Scheller}}, \bibinfo {author} {\bibfnamefont {D.~M.}\ \bibnamefont {Zumbühl}}, \bibinfo {author} {\bibfnamefont {J.~D.}\ \bibnamefont {Zimmerman}},\ and\ \bibinfo {author} {\bibfnamefont {A.~C.}\ \bibnamefont {Gossard}},\ }\bibfield  {title} {\bibinfo {title} {Gaas quantum dot thermometry using direct transport and charge sensing},\ }\href {https://doi.org/10.1007/s10909-014-1169-6} {\bibfield  {journal} {\bibinfo  {journal} {Journal of Low Temperature Physics}\ }\textbf {\bibinfo {volume} {175}},\ \bibinfo {pages} {784} (\bibinfo {year} {2014})}\BibitemShut {NoStop}%
\end{thebibliography}%

\end{document}